%% file: electroweakinos.tex
\pdfoutput=1
\documentstyle[12pt,graphicx,color,cancel,cite,ulem,multirow,framed,hyperref,url,subfigure,amsmath,amssymb]{article}

\def\issue(#1,#2,#3){{\bf #1}, #2 (#3)}

\catcode`\@=11
\def\lsim{\mathrel{\mathpalette\@versim<}}
\def\gsim{\mathrel{\mathpalette\@versim>}}
\def\@versim#1#2{\vcenter{\offinterlineskip
\ialign{$\m@th#1\hfil##\hfil$\crcr#2\crcr\sim\crcr } }}
\catcode`\@=12

\catcode`@=12
\hypersetup{
   bookmarks=true,         % show bookmarks bar?
   unicode=false,          % non-Latin characters in AcrobatÍs bookmarks
   pdftoolbar=true,        % show AcrobatÍs toolbar?
   pdfmenubar=true,        % show AcrobatÍs menu?
   pdffitwindow=false,     % window fit to page when opened
   pdfstartview={FitH},    % fits the width of the page to the window
   pdftitle={My title},    % title
   pdfauthor={Author},     % author
   pdfsubject={Subject},   % subject of the document
   pdfcreator={Creator},   % creator of the document
   pdfproducer={Producer}, % producer of the document
   pdfkeywords={keyword1} {key2} {key3}, % list of keywords
   pdfnewwindow=true,      % links in new window
   colorlinks=true,       % false: boxed links; true: colored links
   linkcolor=blue,        % color of internal links
   citecolor=red,         % color of links to bibliography
   filecolor=magenta,      % color of file links
   urlcolor=cyan,           % color of external links
   linktocpage = true,
   }

\def\beq {\begin{equation}}
\def\eeq {\end{equation}}
\def\bi {\begin{itemize}}
\def\ei {\end{itemize}}
\def\bea {\begin{eqnarray}}
\def\eea {\end{eqnarray}}
\def \PMET{\rm p{\!\!\!/}_T}

\def \met{\rm E{\!\!\!/}_T}

\newcommand{\br}{\begin{eqnarray}}
\newcommand{\er}{\end{eqnarray}}
\newcommand{\be}{\begin{equation}}
\newcommand{\ee}{\end{equation}}

\newcommand{\ch}{\widetilde \chi^\pm}

\def\lum             {{\cal L}}
\newcommand{\ifb} {\rm {fb}^{-1}}

\newcommand{\newc}{\newcommand}
\newc{\wt}{\widetilde}
\newc{\ra}{\rightarrow}

\def \ch2p {{\wt\chi_2^+}}

\def \ch2m {{\wt\chi_2^-}}

\def \chonepm{{\wt\chi_1}^{\pm}}
\def \chonemp{{\wt\chi_1}^{\mp}}
\def \mchonepm{m_{\chonepm}}
\def \mchonemp{m_{\chonemp}}

\newc{\dmchi}{\Delta m_{\wt\chi}}

\def \lspone{\wt\chi_1^0}
\def \mlspone{m_{\lspone}}
\def \lsptwo{\wt\chi_2^0}
\def \mlsptwo{m_{\lsptwo}}
\def \lspthree{\wt\chi_3^0}

\def \slepton{\wt l}
\def \mslepton{m_{\slepton}}

\def\issue(#1,#2,#3){{\bf #1}, #2 (#3)}%AIP format!Vol,page(Year)
\def\iss(#1,#2,#3){{\bf #1} (#3) #2}%AIP format!Vol,page(Year)
\def\ASTR(#1,#2,#3){Astropart.\ Phys. \issue(#1,#2,#3)}
\def\AJ(#1,#2,#3){Astrophysical.\ Jour. \issue(#1,#2,#3)}
\def\AJS(#1,#2,#3){Astrophys.\ J.\ Suppl. \issue(#1,#2,#3)}
\def\APP(#1,#2,#3){Acta.\ Phys.\ Pol. \issue(#1,#2,#3)}
\def\JCAP(#1,#2,#3){Journal\ XX. \issue(#1,#2,#3)} %spdas
\def\SC(#1,#2,#3){Science \issue(#1,#2,#3)}
\def\PRD(#1,#2,#3){Phys.\ Rev.\ D \issue(#1,#2,#3)}
\def\PR(#1,#2,#3){Phys.\ Rev.\ \issue(#1,#2,#3)} % spdas check 
\def\PRC(#1,#2,#3){Phys.\ Rev.\ C \issue(#1,#2,#3)}
\def\NPB(#1,#2,#3){Nucl.\ Phys.\ B \issue(#1,#2,#3)}
\def\NPPS(#1,#2,#3){Nucl.\ Phys.\ Proc. \ Suppl \issue(#1,#2,#3)}
\def\NJP(#1,#2,#3){New.\ J.\ Phys. \issue(#1,#2,#3)}
\def\JP(#1,#2,#3){J.\ Phys.\issue(#1,#2,#3)}
\def\PL(#1,#2,#3){Phys.\ Lett. \issue(#1,#2,#3)}
\def\ZP(#1,#2,#3){Z.\ Phys. \issue(#1,#2,#3)}
\def\ZPC(#1,#2,#3){Z.\ Phys.\ C  \issue(#1,#2,#3)}
\def\PREP(#1,#2,#3){Phys.\ Rep. \issue(#1,#2,#3)}
\def\PRL(#1,#2,#3){Phys.\ Rev.\ Lett. \issue(#1,#2,#3)}
\def\MPL(#1,#2,#3){Mod.\ Phys.\ Lett. \issue(#1,#2,#3)}
\def\RMP(#1,#2,#3){Rev.\ Mod.\ Phys. \issue(#1,#2,#3)}
\def\SJNP(#1,#2,#3){Sov.\ J.\ Nucl.\ Phys. \issue(#1,#2,#3)}
\def\CPC(#1,#2,#3){Comp.\ Phys.\ Comm. \issue(#1,#2,#3)}
\def\IJMPA(#1,#2,#3){Int.\ J.\ Mod. \ Phys.\ A \issue(#1,#2,#3)}
\def\MPLA(#1,#2,#3){Mod.\ Phys.\ Lett.\ A \issue(#1,#2,#3)}
\def\PTP(#1,#2,#3){Prog.\ Theor.\ Phys. \issue(#1,#2,#3)}
\def\RMP(#1,#2,#3){Rev.\ Mod.\ Phys. \issue(#1,#2,#3)}
\def\NIMA(#1,#2,#3){Nucl.\ Instrum.\ Methods \ A \issue(#1,#2,#3)}
\def\EPJC(#1,#2,#3){Eur.\ Phys.\ J.\ C \issue(#1,#2,#3)}
\def\RPP (#1,#2,#3){Rept.\ Prog.\ Phys. \issue(#1,#2,#3)}
\def\PPNP(#1,#2,#3){ Prog.\ Part.\ Nucl.\ Phys. \issue(#1,#2,#3)}
\newc{\PRDR}[3]{{Phys. Rev. D} {\bf #1}, Rapid  Communications, #2 (#3)}

\def\PLB(#1,#2,#3){Phys.\ Lett.\ B  \iss(#1,#2,#3)}
\def\JHEP(#1,#2,#3){JHEP \iss(#1,#2,#3)}
\begin{document}

\begin{flushright}
{HRI-RECAPP-2016-006}
\end{flushright}

\begin{center}

%{\large \bf  Search for Electroweakinos in MSSM with and without Higgs final state  \\ 
% Electroweakinos in pMSSM after LHC Run-1 } \\    

{\large \bf Revisiting the Exclusion Limits from Direct Chargino-Neutralino   \\ 
 Production at the LHC } \\

\vskip 0.5cm
Arghya Choudhury$^{a,b}$\footnote{a.choudhury@sheffield.ac.uk}
Subhadeep Mondal$^{c}$\footnote{subhadeepmondal@hri.res.in}
\vskip 0.3cm

{$^a$Consortium for Fundamental Physics, Department of Physics and Astronomy, \\
University of Sheffield, Sheffield S3 7RH, United Kingdom}

{$^b$Consortium for Fundamental Physics, Department of Physics and Astronomy, \\
University of Manchester, Manchester, M13 9PL, United Kingdom}

{$^c$ Regional Centre for Accelerator-based Particle Physics,\\
Harish-Chandra Research Institute, Jhusi, Allahabad - 211019, India}

\end{center}

%\pacs{}

\vskip 0.3cm 
%%%%%%%%%%%%%%%%%%%%%%%
\begin{abstract}
%%%%%%%%%%%%%%%%%%%%%%%
We revisit the existing limits on the gaugino masses in various Supersymmetric 
(SUSY) scenarios derived from Run-I data of the LHC. These limits obtained from 
the various final states rely heavily on the simplified assumptions regarding the 
masses, compositions and decay branching ratios of the gauginos.  
The most severe exclusion limits on the gaugino masses are obtained 
from trilepton final states while the second lightest neutralino ($\widetilde \chi_2^0$) 
decaying into the SM-like Higgs and lightest SUSY particle (LSP) results in the weakest  
bounds. Our aim is to assess the extent of deviation of these 
exclusion limits in more realistic scenarios. After a brief discussion 
on the various decay modes of the $\widetilde \chi_2^0$ and 
the lightest chargino ($\widetilde\chi^{\pm}_1$), we proceed to validate the ATLAS exclusion 
limits obtained from trilepton, $l\gamma\gamma$ and $lb\bar b$ final 
states associated with missing energy. We then consider different combinations of the 
relevant branching ratios to study their impact on the existing bounds. 
The results are presented alongside the existing exclusion limits to showcase the 
extent of the obtained deviation. We also observe that the three-body decay modes of 
$\widetilde\chi_2^0$ and $\widetilde\chi^{\pm}_1$ via off-shell slepton decays resulting in 
trilepton final states provide bounds that are far more severe in some parts of 
the available parameter space than that obtained from the off-shell gauge boson decays.  
%%%%%%%%%%%%%%%%%%%%%%
\end{abstract}
%%%%%%%%%%%%%%%%%%%%%%
\newpage
\setcounter{footnote}{0}
%%%%%%%%%%%%%%%%%%%%%% Table of content %%%%%%%
%\hypertarget{toc}{}
%\small
\hrule
\tableofcontents
\vskip 1.0cm
\hrule
%%%%%%%%%%%%%%%%%%%%%%%%%%%%%%%%%%%%%%%%%%%%%%%
%%%%%%%%%%%%%%%%%%%%%%%%%%%%%%%%%%%%%%%%%%%%%%%
%\newpage 
%%%%%%%%%%%%%%%%%%%%%%%%%%%%%%%
\section{Introduction}
%%%%%%%%%%%%%%%%%%%%%%%%%%%%%%%
After a long shutdown, the Large Hadron collider (LHC) is now operating in 
full swing at a center of mass energy, $\sqrt s$ = 13 TeV. In the aftermath 
of its huge success in terms of the discovery of the elusive Higgs boson 
with mass around 125 GeV\cite{atlas_125,cms_125} , 
 the prime goal of Run-II now is to look for new physics beyond the 
 Standard Model (BSM). So far, after analysing the Run-I data, 
ATLAS and CMS have only reported some small inconclusive local excesses 
\cite{8tev_excess1,8tev_excess2} 
over the SM predictions, which need to be put under thorough scrutiny at Run-II. 
Any of these excesses, if proven significant, will open the window to the 
hitherto unknown BSM physics. On the other hand, after just a few months of data 
accumulation at $\sqrt s$ = 13 TeV, both CMS and ATLAS have hinted towards a 
possible scalar resonance at 750 GeV \cite{ATLAS-run-II-1, CMS-run-II-2} 
that has created a lot of buzz within the particle physics community. 
Although promising, one has to wait for more data and finer analyses 
to ascertain if this is indeed the first hint of the BSM physics we are so eagerly 
waiting for. At this stage, it is therefore worthwhile to look back to our favourite 
new physics models and revisit the exclusion limits derived from the existing 
experimental data. Supersymmetry (SUSY) \cite{SUSYreviews1,SUSYreviews2,SUSYbooks}, 
being one the frontrunners among the BSM candidates, has been searched for extensively 
both by the ATLAS and the CMS. The experimental results so far indicate towards a 
heavy coloured sector {\it(strongly interacting  sector)} within the framework of 
the minimal supersymmetric SM (MSSM)\cite{atlas_old,atlas_new,cmsall}. 
ATLAS and CMS have already excluded first two generation squarks and gluino 
masses upto 1.7 TeV for degenerate squark-gluino scenario 
\cite{atlas_old,atlas_new,cmsall} \footnote{However, these strong bounds reduce 
significantly in compressed SUSY type scenarios \cite{compressed}.} from Run-I data. 
Depending upon the scenarios, the very recent Run-II data with an integrated 
luminosity ($\lum$) = 3.2 $\ifb$, the limit on gluino and squarks 
has already been increased by 100 to 200 GeV \cite{atlas13,cms13}.

At the same time, exclusion limits on the masses of the electroweak sector 
sparticles, namely, charginos, neutralinos \footnote{In this paper 
we will refer to the charginos and neutralinos as the electroweakinos.} and 
sleptons, are {\it much} weaker 
because of their relatively smaller production cross-sections at the LHC. 
The stringent limits on the coloured sector sparticles make 
the search of the electroweakinos and the sleptons all the more important 
from the perspective of SUSY searches at the 13 and 14 TeV run of the LHC. 
As the ATLAS and the CMS collaborations have not yet published any new search results 
of the electroweak sparticles from the very first 13 TeV 3.2 $\ifb$ data, here we will 
only focus on the Run-I data and the corresponding limits. From the combined data 
obtained at 7 TeV and 8 TeV runs, the experimental collaborations have already 
put significant mass limits on the electroweakinos and the sleptons from direct 
search channels \cite{atlas_ew_3l,atlas_ew_2l,atlas_ew_tau,atlas_ew_higgs, 
atlas_ew_summary,cmsew1,cmsew2} that include chargino - neutralino, chargino pair 
and slepton pair productions. However, the existing bounds apply only for 
some simplified SUSY scenarios which assume a certain hierarchy between the 
slepton and the gaugino masses and  also fix their compositions. In the 
studies of the electroweakinos, for example, the lightest neutralino ($\lspone$) 
which is also assumed to be the lightest supersymmetric particle (LSP) is usually 
supposed to be purely bino-like while the second lightest neutralino 
($\lsptwo$) and the lighter chargino ($\chonepm$) are purely wino-like 
\cite{atlas_ew_3l,atlas_ew_2l,atlas_ew_tau,atlas_ew_higgs, 
atlas_ew_summary,cmsew1,cmsew2}. These assumptions have crucial impact on the 
relevant production cross-section and decay branching ratio (BR) of 
the concerned sparticle and as a result, the existing bounds are expected to change 
significantly in more realistic scenarios. 

The most stringent bound on chargino - neutralino mass plane is obtained 
from $\chonepm \lsptwo$ pair production leading to various trilepton + transverse 
missing energy ($\met$) final states \cite{atlas_ew_3l,cmsew2}. 
A number of phenomenological studies already exist which have 
analysed the implications of electroweakino searches and related topics 
at the LHC \cite{electroweakino,ucmc1,ucmc2,spoiler,spoiler_Bharucha}. 
In a recent work \cite{ucmc2} it has been shown that the LHC constraints in 
the trilepton channel are significantly weaker even in the presence of light 
sleptons ($\mslepton < m_\chonepm$), especially in the models with higgsino 
dominated $\chonepm, \lsptwo$ and $\lspthree$; compared to the scenarios 
mostly studied by the LHC collaborations with wino-dominated $\chonepm, \lsptwo$. 
For such scenarios with higgsino dominated gauginos 
(or gauginos with non-negligible higgsino component)  
Br($\tilde \chi_3^{0}, \tilde \chi_2^{0} \rightarrow h \tilde\chi_1^{0}$) is 
usually large and as a consequence reduces the signal significance of the trilepton 
final states. Even for wino-dominated $\chonepm$ and $\lsptwo$ and with decoupled 
sleptons, the limits obtained by ATLAS and CMS are much weaker when 
$\lsptwo$ decays via  $Z \lspone$. One obtains the weakest exclusion limits when 
$\lsptwo$ decays via the {\it spoiler} mode ($h \lspone$) 
\footnote{A few phenomenological analyses in this context may be seen in 
Ref. \cite{ucmc2,spoiler,spoiler_Bharucha}.}  \cite{atlas_ew_higgs, atlas_ew_summary,cmsew2}. 
In the decoupled - sleptons scenarios, the subsequent two body decays 
$\chonepm\rightarrow\lspone W^{\pm}$ and $\lsptwo\rightarrow\lspone Z/h$ result 
in various leptonic final states associated with b-jets or photons and $\met$. 
ATLAS collaboration has recently presented the electroweakino search results with 
$1l + 2b$ and $1l + 2\gamma$ final states \cite{atlas_ew_higgs}. 
However, the limits are derived on the gaugino mass plane assuming 
100$\%$ branching ratio into one particular decay mode, which does not provide 
us with the whole picture. 
The interplay between the two-body decay modes like 
($\lsptwo \to h \lspone$ and $\lsptwo \to Z\lspone $) can also modify the 
exclusion limits considerably in more realistic situations, 
e.g., phenomenological MSSM (pMSSM) \cite{pmssm}. It is, therefore, worthwhile to revisit 
the existing limits with different compositions and mass hierarchies 
of the gauginos. Moreover, the off-shell production and subsequent decays of 
the gauge bosons and sleptons from $\lsptwo / \chonepm$ decays can result in 
a wide range of possible decay BRs. These off-shell decay modes, if taken 
into account, can also give rise to similar final states as obtained from 
the two-body decay modes of the gauginos. All these possibilities need to be 
explored in the light of the data accumulated from run-I before embarking to 
the run-II of the LHC. 

A light gaugino and (or) slepton scenario is also highly motivated 
from the observed excess of anomalous magnetic moment of muon $((g-2)_{\mu})$ 
measurements and existence of Dark Matter (DM). 
In order to fit the excess in $(g-2)_{\mu}$ \cite{g-2exp} over the SM predictions 
within the framework of MSSM, requires the slepton and the lighter chargino 
masses in the few hundreds of GeV range \cite{ucmc1,ucmc2,jcacsm}. 
Various leptonic final states associated with large missing energy 
are the favoured channels to look for such a scenario. A heavy coloured sector, 
as favoured by the LHC data, fits more naturally for such MSSM parameter space as 
their contributions to the cascade of the gauginos from even off-shell productions 
are small in that case resulting in larger branching ratios to the 
leptonic final states resulting in cleaner signals.   In all R-parity conserving 
SUSY models, the LSP (in our case, $\lspone$)  is stable and can be a good 
candidate for the elusive dark matter (DM) \cite{dm_rev1,silk, dm_rev2} 
in the universe. 
A partial list of works on supersymmetric DM may be seen in  Ref.~\cite{dmmany,dmsugra,dmmssm,dmmssmfurther,dmmssmnew,Han:2013gba}. 
The electroweak sparticles may also lead to correct relic density for DM 
via the DM annihilation/coannihilation mechanisms.
This way the electroweak sector can also be constrained from the  precisely 
measured value of DM relic density by WMAP \cite{wmap} or PLANCK \cite{planck}. 

In this present study, we assume that the sleptons are heavier than the electroweakinos, 
but not so heavy that they may be considered to be decoupled. The entire coloured 
sector, on the other hand, is decoupled from the rest of the SUSY spectrum. 
%The current search strategies by ATLAS and CMS for slepton pair production are 
%based on the assumption that sleptons are next lightest supersymmetric particle 
%(NLSP) and BR($\slep \to l \lspone$ = 100\%). 
Apart from the obvious advantages of having light sleptons in the theory from the 
viewpoint of $(g-2)_{\mu}$ and DM, such scenarios 
may result in significant enhancement of the three body branching 
ratios of the electroweakinos from the off-shell decays of the sleptons. 
Such off-shell slepton decays in this context of the gauginos have not been 
studied by the experimental collaborations so far. 
%resulting in significant modifications in the exclusion limits imposed by 
%ATLAS and CMS  when the elctroweakinos decay via virtual W or Z.   
%Hence it is worthwhile to study the possibilities of light electroweakinos 
%with or without light sleptons considering the current LHC, $(g-2)_{\mu}$ and 
%DM relic density data. 

The paper is organised in the following way. First, we explore the available 
parameter space in pMSSM through a detailed scan using the constraints derived 
from the most updated collider and flavour physics data. 
In Sec.~\ref{sec2} we discuss about the various decay modes of 
$\lsptwo$ and $\chonepm$ with 
different slepton-gaugino mass hierarchies to study the variation of 
their various 2-body and 3-body decay branching ratios. 
Here we also briefly discuss about the impact of muon g-2 and DM relic 
density on the available parameter space.
In Sec.~\ref{sec3}, we study the direct pair production of $\lsptwo\chonepm$ 
and their subsequent decays resulting in various final states following 
the footsteps of the corresponding ATLAS analyses for validation. 
Then we proceed with the same scenarios with different values of the  
2-body and 3-body branching ratios of $\lsptwo$ and $\chonepm$ to 
produce similar final states and revisit the present exclusion limits 
imposed by the ATLAS analyses. We finally provide our conclusions 
in Sec.~\ref{sec4}.       
%%%%%%%%%%%%%%%%%%%%%%%%%%%%%%%
\section{Probing the Electroweak Sector}
\label{sec2}
%%%%%%%%%%%%%%%%%%%%%%%%%%%%%%%
In this section, first we briefly discuss about the various two-body and 
three-body decay modes of $\lsptwo$ and $\chonepm$ to study the interplay among 
the branching ratios for different choices of gaugino compositions and 
slepton - gaugino mass hierarchies. In the simplified model scenarios,
the exclusion limits on the gaugino masses are derived assuming 
both $\lsptwo$ and $\chonepm$ decay into any one of their respective 
available two-body decay modes with 100\% branching ratio. 
Our aim is to find combinations of different decay modes of these particles 
that also may give rise to similar final states. Therefore, in order to 
obtain a clear idea of the variation of the branching ratios of 
$\lsptwo$ and $\chonepm$ over the pMSSM parameter space, we scan the relevant parameters 
in the following ranges: \\ 
\begin{eqnarray}
1~{\rm GeV} < M_1 < 1200~{\rm GeV},&\ \  100 ~ {\rm GeV} <M_2 < 1500~ {\rm GeV},&\ \ 100 ~ {\rm GeV} <\mu < 2000~ {\rm GeV}  \nonumber \\
1 < \tan\beta < 50,&\ \  100~ {\rm GeV} < M_{l_L} < 2000~ {\rm GeV}, &\ \  100~ {\rm GeV} < M_{l_R} < 2000~ {\rm GeV},  \nonumber
\label{parameterRanges}
\end{eqnarray}
where, $M_1$, $M_2$ and $\mu$ are the bino, wino and higgsino soft mass parameters respectively.  
$\tan\beta$ is the ratio of the up-type and down-type Higgs vacuum expectation values. 
$M_{l_L}$ and $M_{l_R}$ denote the left and right-handed slepton 
soft masses respectively. We assume equal soft masses for all three 
generations sleptons. The gluino and the squark sector particles have 
no impact in our present study. Hence we decouple these particles from the rest 
of the spectrum and keep their soft mass parameters at 3 TeV.
For this scan we have used SUSPECT\cite{suspect} and SUSY-HIT\cite{susyhit} to calculate the 
SUSY spectrum and the relevant branching ratios. The flavour and other 
low energy constraints have been calculated using micrOMEGAs\cite{micromega3}. 
While scanning we ensure that 
for all the points $M_1 < M_2 < \mu$, so that the LSP is always mostly bino-like. However, 
since $M_1$, $M_2$ and $\mu$ have all been varied independent of each other, 
the LSP may as well be a bino-wino or bino-wino-higgsino mixed state.  
We also make sure that for all our points, $\chonepm$ (mostly wino-like) is the NLSP.
As a result of our choice of the gaugino mass parameters, the $\chonepm$ is either 
wino-like or a wino-higgsino mixed state. $\lsptwo$ is also mostly expected to 
be wino-like and have a mass close to that of $\chonepm$. However, it may also 
be a wino-bino, wino-higgsino or wino-bino-higgsino mixed state. As a 
consequence of having $\chonepm$ (and $\lsptwo$ if these masses are degenerate), 
as NLSP, none of $\lsptwo$ or $\chonepm$ can decay into an on-shell slepton or sneutrino. 
However, these sfermions can be produced off-shell and the three 
body decay modes of the electroweakinos may have large enough branching 
ratios which can not be ignored. We discuss more about this later in 
this section. The following experimental constraints have been taken 
into account while scanning: 
\begin{itemize}
\item The lightest CP-even Higgs boson mass should be in the 
range 125 $\pm 3$ GeV \cite{atlas_125,cms_125} considering a theoretical 
uncertainty of 3 GeV \cite{higgsuncertainty3GeV}.
\item Lighter chargino ($\chonepm$) mass should be above LEP 
exclusion limit, i.e, 103.5 GeV \cite{lepsusy}.  
\item We impose the flavour physics constraints: \\
$2.82 \times 10 ^{-4} < $ BR($b \ra s \gamma$) $ < 4.04 \times 10 ^{-4}$ (at 2$\sigma$ level) \cite{Amhis:2014hma} and \\
$1.57 \times 10 ^{-9} < $ BR($B_s \ra \mu^+ \mu^-$) $ < 4.63 \times 10 ^{-9}$ (at 2$\sigma$ level) \cite{Amhis:2014hma}.
\item Limits on slepton masses: Both ATLAS and CMS have looked for sleptons 
via direct production channels with di-lepton final states associated with $\met$ \cite{atlas_ew_2l}. 
In these analyses sleptons are considered as NLSP and BR($\slepton \to l \lspone$) = 100\%. 
For example, with degenerate Left(L) and Right(R) type slepton masses, 
LHC data exclude the region $90 < m_{\wt l} < 325$ GeV for a massless LSP. 
As the LSP-slepton mass splitting decreases, the exclusion limit becomes weaker. 
Again only for L-type or R-type slepton production, the exclusion limit is 
relatively weaker than the degenerate case (see Fig. 8 of Ref.~\cite{atlas_ew_2l}). 
In our parameter space scan, $\chonepm$ and $\lsptwo$ are always assumed to be 
lighter than the sleptons. In most of the regions, R-type sleptons dominantly 
decay to $l \lspone$, but the L-type sleptons can have significant branching ratio in 
the additional decay modes like $\nu \chonepm$ or $l \lsptwo $ which have not been 
considered by the LHC collaborations. In such scenarios, limits on the slepton (L-type) 
masses can change from the existing limits. A detailed computation of these revised 
limits is beyond the scope of this work. Instead, we have implemented the bounds on 
the slepton masses in a bin-by-bin basis separately for L-type and 
R-type sleptons derived from Fig. 8 of Ref.~\cite{atlas_ew_2l} depending on various slepton and 
LSP mass regions.        
\end{itemize}
%%%%%%%%%%%%%%%%%%%%%%%%%%%%%%
\subsection{$\lsptwo$ decay modes}
%%%%%%%%%%%%%%%%%%%%%%%%%%%%%
The dominant two-body decay modes of the second lightest 
neutralino ($\lsptwo$) are $\lsptwo\rightarrow\lspone h$, 
$\lsptwo\rightarrow\lspone Z$ and $\lsptwo\rightarrow f \widetilde f$, 
where $h$ denotes the SM-like lightest CP even Higgs boson\footnote{We 
assume all the other Higgs bosons in the MSSM are decoupled 
from rest of the spectrum.} and $f (\widetilde f)$ denotes the 
fermions (sfermions). 
If the squarks and sleptons in the theory are heavier 
than $\lsptwo$, depending on the mass difference between $\lsptwo$ and 
$\lspone$ ($\Delta m^0_{\wt\chi}$), any one of the other two decay 
modes dominate or compete with each other. 
%%%%%%%%%%%%%%%%%%%%%%%%%%%%%%%
\begin{figure}[h!]
\centering
\hspace{-1cm}
\includegraphics[angle =0, width=0.48\textwidth] {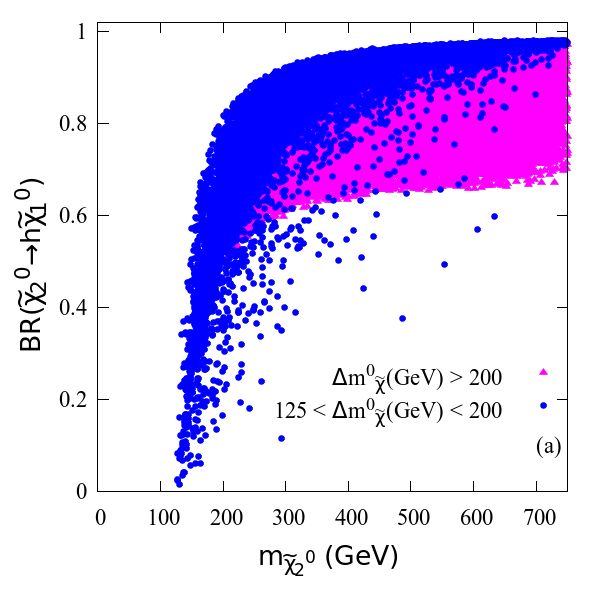}.
\includegraphics[angle =0, width=0.48\textwidth] {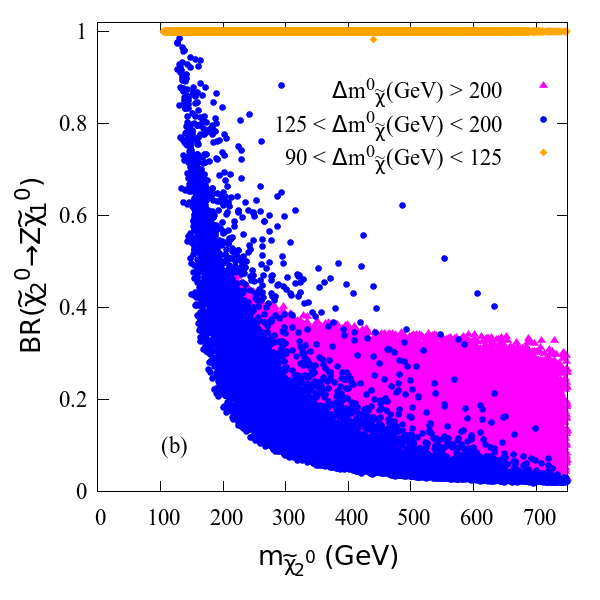}
\caption{Distributions of the BR's corresponding to the two 
 2-body decay modes of $\lsptwo$ shown as 
a function of $m_{\lsptwo}$. The left plot (a) shows the distributions 
of BR($\lsptwo\rightarrow h\lspone$) and 
the right plot (b), that of BR($\lsptwo\rightarrow Z\lspone$). 
The different coloured points in the plots 
correspond to the different $\Delta m^0_{\wt\chi} = \mlsptwo - \mlspone$ as 
indicated in the plot.}
\label{fig:neut2-2body}
\end{figure}
%%%%%%%%%%%%%%%%%%%%%%%%%%%%%%%

In Fig.~\ref{fig:neut2-2body}, we show the distribution of the 
branching ratios of the two 2-body decay 
modes of the $\lsptwo$ into $h$ and $Z$ final states associated with $\lspone$. 
The distributions are shown as a function of the $\lsptwo$ mass ($\mlsptwo$). 
Effect of the mass difference, $\Delta m^0_{\wt\chi}$ on the 
BR of these decay modes can be understood by the different colours 
and shapes of the points. We take different mass windows of $\Delta m^0_{\wt\chi}$ to 
showcase the effect. The magenta points correspond to 
$\Delta m^0_{\wt\chi} > 200$ GeV. For such a large mass difference, 
both the decay modes are open. However, in this case, 
BR($\lsptwo\rightarrow h\lspone$) $>$ BR($\lsptwo\rightarrow Z\lspone$). 
This is due to the fact that the $Z$ boson only couples 
with the neutralinos via their higgsino components. 
Higgsino components in both the $\lspone$ and the $\lsptwo$ 
states being small in most of the points, this decay mode 
is generally suppressed compared to the other. The blue circle points 
correspond to $125 < \Delta m^0_{\wt\chi} < 200$ GeV. 
Again we obtain similar pattern except for the points where 
the mass difference is barely sufficient to produce $h$ in 
the final state. Once there is enough phase space for the decay, 
$\lsptwo\rightarrow h\lspone$ to take place, its BR starts to 
dominate. Orange triangle points in Fig.~\ref{fig:neut2-2body}b
correspond to $90 < \Delta m^0_{\wt\chi} < 125$ GeV. Naturally, the 
$\lsptwo$ decays entirely to $Z\lspone$ mode as the other decay mode 
is  kinematically inaccessible.

%%%%%%%%%%%%%%%%%%%%%%%%%%%%%%%
\begin{figure}[!h]
\includegraphics[angle =0, width=0.32\textwidth] {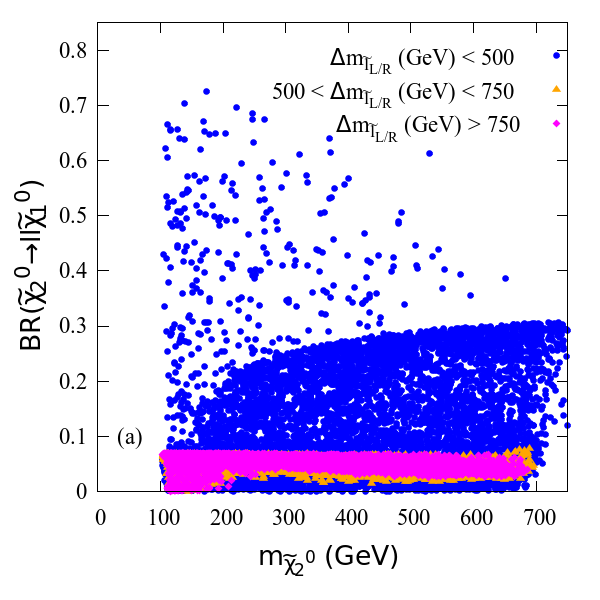}
\includegraphics[angle =0, width=0.32\textwidth] {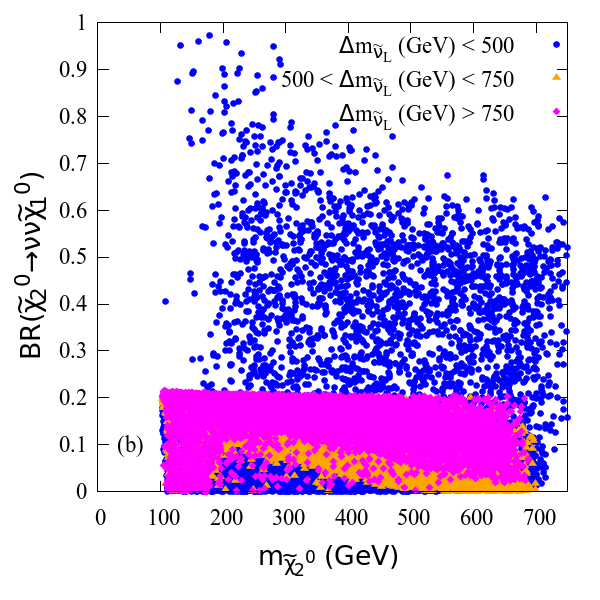}
\includegraphics[angle =0, width=0.32\textwidth] {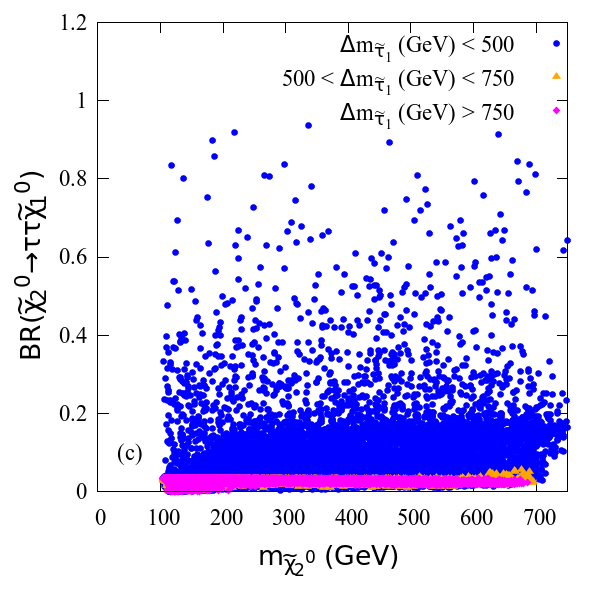}
\caption{Distributions of the BR's corresponding to the three 
leptonic 3-body decay modes of $\lsptwo$ shown as 
a function of $m_{\lsptwo}$. The plots show the distributions 
corresponding to BR($\lsptwo\rightarrow l\bar l\lspone$) in left (a), 
BR($\lsptwo\rightarrow\nu\bar\nu\lspone$) in middle (b) 
and BR($\lsptwo\rightarrow\tau\bar\tau\lspone$) in right (c) respectively. 
Note that, BR($\lsptwo\rightarrow l\bar l\lspone$) contains contributions 
from both the electron and muon associated final states 
while BR($\lsptwo\rightarrow\nu\bar\nu\lspone$) contains all the 
three neutrino flavour contributions. The different 
coloured (shaped) points correspond to the different 
%$\Delta m_{\wt l_{L/R}}$.
$\Delta m_{\wt l_{L/R}/{\wt\nu_L}/{\wt\tau_1}} = m_{\wt l_{L/R}/{\wt\nu_L}/{\wt\tau_1}} - 
 m_{\lsptwo}$ (see text for details).}
\label{fig:neut2-3body}
\end{figure}
%%%%%%%%%%%%%%%%%%%%%%%%%%%%%%

Fig.~\ref{fig:neut2-3body} shows the distribution of the BRs of the 
three 3-body decay modes of $\lsptwo$ into different 
leptonic final states. Once the two-body decay modes become 
kinematically inaccessible, these three-body decays start 
to show up. The contribution to these decay modes may come 
from off-shell sleptons as well as off-shell $Z$ or $h$ decays. 
To determine how heavy sleptons affect these decays, we have 
plotted the BRs as a function of $\mlsptwo$ corresponding to different mass ranges of 
$\Delta m_{\wt l_{L/R}/{\wt\nu_L}/{\wt\tau_1}} = m_{\wt l_{L/R}/{\wt\nu_L}/{\wt\tau_1}} - 
 m_{\lsptwo}$ denoting them by different colored points. 
While plotting BR($\lsptwo\rightarrow l\bar l\lspone$), 
the $\wt l$ - $\lsptwo$ mass gap is calculated by choosing the smaller 
mass between $m_{\wt l_{L}}$ or $m_{\wt l_{R}}$ as the slepton 
mass since both the L-type or the R-type sleptons may affect the 
branching ratio.  Note that, BR($\lsptwo\rightarrow l\bar l\lspone$) 
includes both electron and muon final states. However, while plotting 
 BR($\lsptwo\rightarrow\nu\bar\nu\lspone$), we only consider different 
ranges of $\Delta m_{\wt\nu_L}$. BR($\lsptwo\rightarrow\nu\bar\nu\lspone$) 
includes sum of all the three neutrino decay modes.    
For BR($\lsptwo\rightarrow\tau\bar\tau\lspone$), we consider $\Delta m_{\wt\tau_1}$. 
As seen from the plots, off-shell slepton decays contribute mostly while 
$\Delta m_{\wt l_{L/R}} (/\Delta m_{\wt\nu_L}/\Delta m_{\wt\tau_1}) < 500$ 
GeV as shown by the blue (circle) points. 
Naturally, we obtain highest BR for the $\tau\bar\tau$ final state, 
stau being the lightest slepton. As the slepton - $\lsptwo$ mass 
difference keeps increasing, contributions from off-shell sleptons 
start to diminish and that from off-shell $Z$-boson start to dominate. 
The Z-boson, despite being much lighter than the sleptons, only starts 
to affect the BRs in cases where the slepton masses are quite heavy 
since for most of the points both $\lspone$ and $\lsptwo$ has small higgsino components and hence their coupling to $Z$-boson is usually suppressed. 
Contribution of the off-shell Higgs boson state is small due to 
its small coupling with the SM leptons. 
The orange (triangle) points correspond to 
$500~{\rm GeV} < \Delta m_{\wt l_{L/R}} (/\Delta m_{\wt\nu_L}/\Delta m_{\wt\tau_1}) < 750~{\rm GeV}$ 
and the magenta (diamond) points correspond to 
a mass difference $> 750$ GeV. 
Note that, the spread in the orange and the magenta points are obtained 
due to the off-shell slepton contributions and the varying higgsino components in the neutralinos.
Looking at the values of the BRs from the plots it 
can be easily understood that once the mass difference becomes greater 
than 500 GeV, the three body decays are entirely controlled by off-shell $Z$-decays.   
One should note that, there exist a large part of the parameter space where $\lsptwo$ 
has a very large branching ratio into the invisible mode 
($\nu\nu\lspone$). None of the usual search channels 
are sensitive to probe such a scenario from $\chonepm$-$\lsptwo$ pair production.  

%%%%%%%%%%%%%%%%%%%%%%%%%%%%%%%
\subsubsection{Impact of sign of $\mu$}
%%%%%%%%%%%%%%%%%%%%%%%%%%%%%%%
Note that, so far we have only concentrated on a positive $\mu$ while deriving our results. 
However, reversing the sign of $\mu$ may alter the results significantly. It is well known 
that the decay width of $\lsptwo\to\lspone h$ depends on the sign of $\mu$ and may go down 
considerably. Under certain approximations in the Higgs decoupling limit, it can be shown 
that this decay branching ratio is proportional to a fator $(\frac{M_1+M_2}{\mu} + 
\frac{4}{{\rm tan}\beta})^2$ \cite{spoiler_Bharucha}. Hence for a negative $\mu$ and $|\mu| >> M_1,M_2$,  
one would expect BR($\lsptwo\rightarrow h\lspone$) 
to be suppressed than BR($\lsptwo\rightarrow Z\lspone$) in that region of parameter space, where 
both the decay modes are open. The cancellation is even more severe as tan$\beta$ decreases. 
To showcase this behavior, we show in Fig.~\ref{fig:neut2-2body-pnmu}, the relative strengths 
of the two concerned decay BR's as a function of $m_{\lsptwo}$.   
For this purpose, we keep the LSP bino-like with $M_1$ fixed at 100 GeV. Lower limit of $M_2$ is chosen 
such that both the decay modes, $\lsptwo\rightarrow Z(h)\lspone$ are kinematically possible.  
All the sleptons and the squarks in the theory are decoupled from rest of the spectrum. $\mu$-value is kept 
fixed at 1 TeV while its sign is varied. To showcase the tan$\beta$ dependence, we choose to present our results 
at two tan$\beta$ values, 10 and 30.
%%%%%%%%%%%%%%%%%%%%%%%%%%%%%%%
\begin{figure}[h!]
\centering
\hspace{-1cm}
\includegraphics[angle =0, width=0.48\textwidth] {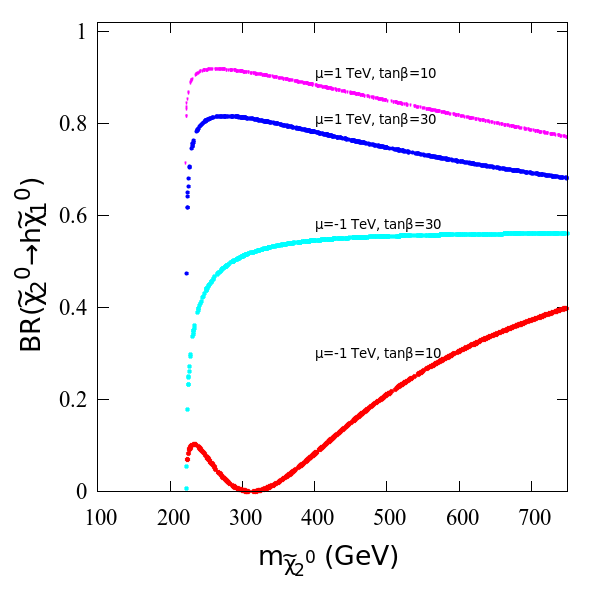}.
\includegraphics[angle =0, width=0.48\textwidth] {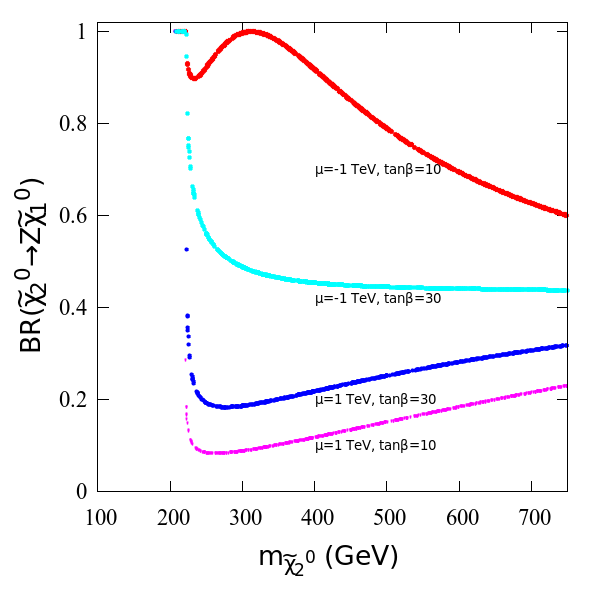}
\caption{Distributions of the BR's corresponding to the two 
 2-body decay modes of $\lsptwo$ shown as a function of $m_{\lsptwo}$ for different 
sign of $\mu$ at a fixed LSP mass and two different tan$\beta$ values. 
The left plot (a) shows the distributions 
of BR($\lsptwo\rightarrow h\lspone$) and 
the right plot (b), that of BR($\lsptwo\rightarrow Z\lspone$). $M_1$ is kept fixed at 100 GeV while 
all the sleptons and the squarks in the theory are decoupled.}
\label{fig:neut2-2body-pnmu}
\end{figure}
%%%%%%%%%%%%%%%%%%%%%%%%%%%%%%%

In Fig.~\ref{fig:neut2-2body-pnmu}, the magenta and blue coloured points correspond to positive $\mu$ for 
tan$\beta$= 10 and 30 respectively. In this case, the distribution of the BR's are similar as already 
depicted in Fig.~\ref{fig:neut2-2body}. On the other hand, red and cyan points corresponds to negative $\mu$ for 
tan$\beta$= 10 and 30 respectively. As expected, for this case, BR($\lsptwo\rightarrow Z\lspone$) dominates 
over BR($\lsptwo\rightarrow h\lspone$) even when both the decay modes are kinematically accessible. Since 
the magnitude of the $\mu$-parameter ($|\mu|$) remains fixed at 1 TeV, the dominance of BR($\lsptwo\rightarrow Z\lspone$) 
over BR($\lsptwo\rightarrow h\lspone$) becomes less prominent as $M_2$ increases.

%%%%%%%%%%%%%%%%%%%%%%%%%%%%%
\subsection{$\chonepm$ decay modes}
%%%%%%%%%%%%%%%%%%%%%%%%%%%%%

The sleptons in the theory being heavier than $\chonepm$, 
it has only one two body decay mode, 
$\chonepm\rightarrow W^{\pm}\lspone$. Therefore, if kinematically allowed, 
BR to this decay mode stands at 100$\%$.
Once $m_{\chonepm} - m_{\lspone} < m_W$, 
the three body decay modes starts to open up.  

%%%%%%%%%%%%%%%%%%%%%%%%%%%%%%%
\begin{figure}[h!]
\centering
\hspace{-1cm}
\includegraphics[angle =0, width=0.48\textwidth] {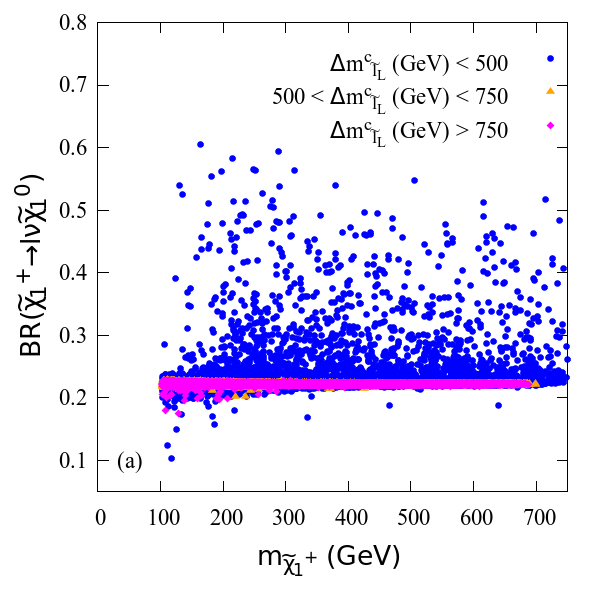}
\includegraphics[angle =0, width=0.48\textwidth] {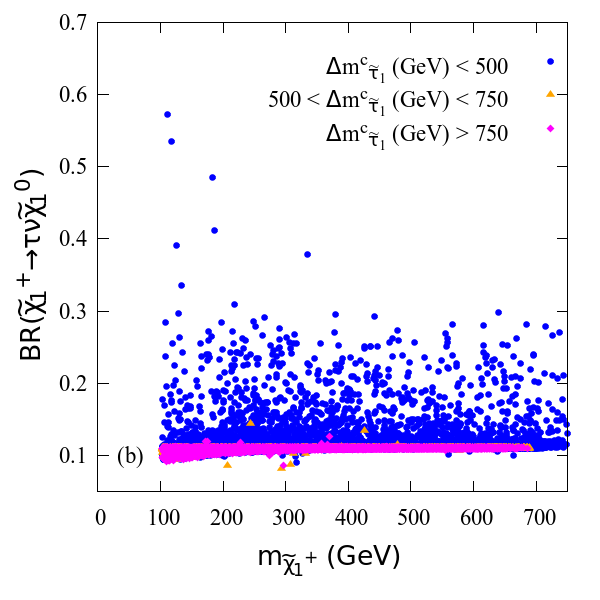}
\caption{Distributions of the BR's corresponding to the 
leptonic 3-body decay modes of $\chonepm$ shown as 
a function of $m_{\chonepm}$. The left plot (a) show the 
distributions corresponding to BR($\chonepm\rightarrow l\nu_l l\lspone$), 
and the right plot (b), that of BR($\chonepm\rightarrow\tau\nu_{\tau}\lspone$). 
Note that, BR($\chonepm\rightarrow l\nu_l l\lspone$) 
contains contributions from both the electron and muon associated 
final states. The different coloured (shaped) points correspond to the different 
$\Delta m^c_{\wt l_{L}/{\wt\tau_1}} = m_{\wt l_L/{\wt\tau_1}} - m_{\chonepm}$.}
\label{fig:char-3body}
\end{figure}
%%%%%%%%%%%%%%%%%%%%%%%%%%%%%%

Fig.~\ref{fig:char-3body} shows the distribution of the BRs of the three 
body decay modes of $\chonepm$ into different 
leptonic final states. Since our $\chonepm$ is mostly wino-like, it 
only couples with the left-handed sleptons. Hence, 
we showcase the effects of the slepton mass on this decay BRs by the 
different coloured points in the plot corresponding 
to different mass ranges of $\Delta m^c_{\wt l_L/{\wt\tau_1}} = m_{\wt l_L/{\wt\tau_1}} - m_{\chonepm}$. 
Apart from the off-shell sleptons, off-shell $W$-bosons may also 
contribute here. However, as in the case of $\lsptwo$, here also
the off-shell sleptons contribute mostly toward the three-body 
decays unless they are too heavy. We plot BR($\chonepm\rightarrow l\nu_l l\lspone$) 
as a summed up contribution of the both electron 
and muon associated decay modes. BR($\chonepm\rightarrow\tau\nu_{\tau}\lspone$) 
dominates as long as the sleptons are relatively 
light ($\Delta m^c_{\wt l_L} (\Delta m^c_{\wt\tau_1}) < 500$ GeV), 
shown as the blue (circle) points in the distributions. 
Once the sleptons start to get heavy, the off-shell $W$-boson 
starts to contribute as shown by the orange (triangle) and magenta (diamond) points, 
corresponding to $500 < \Delta m^c_{\wt l_L} (\Delta m^c_{\wt\tau_1}) < 750$ GeV 
and $\Delta m^c_{\wt l_L} (\Delta m^c_{\wt\tau_1}) > 750$ GeV respectively. 

%%%%%%%%%%%%%%%%%%%%%%%%%%%%%%%
\begin{figure}[h!]
\centering
\hspace{-1cm}
\includegraphics[angle =0, width=0.6\textwidth] {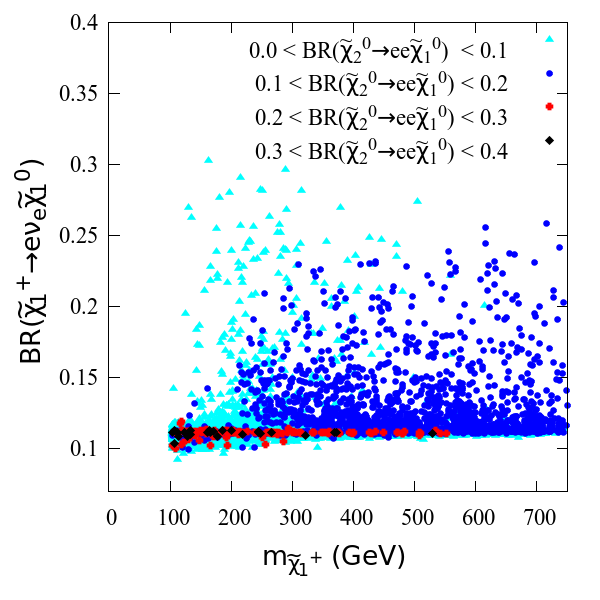}
\caption{Distribution of BR($\chonepm\rightarrow\lspone e^{\pm}\nu_{e}$) shown as a function of $m_{\chonepm}$. 
The cyan, blue, red and black points indicate BR($\lsptwo\rightarrow\lspone e^{\pm}e^{\mp}$) $<$ 10$\%$, 
10$\%$-20$\%$, 20$\%$-30$\%$ and 30$\%$-40$\%$ respectively. }
\label{fig:3bd_char_neut}
\end{figure}
%%%%%%%%%%%%%%%%%%%%%%%%%%%%%%

The large three-body 
branching ratios of the gauginos in part of the parameter space indicate 
that comparable exclusion limits may be derived 
if one considers them as well instead of considering only the two-body 
decay modes. To emphasise this point, as an example, we show 
in Fig~\ref{fig:3bd_char_neut}, the  distribution of 
BR($\chonepm\rightarrow\lspone e^{\pm}\nu_{e}$) as a function of $m_{\chonepm}$. 
The different colour codes indicate different regions of the 3-body  
decay BR($\lsptwo\rightarrow\lspone e^{\pm}e^{\mp}$).  

This plot gives an idea of the relative abundance of the three-body decay modes. 
Note that if one demands a large BR for the decay mode $\lsptwo\rightarrow\lspone e^{\pm}e^{\mp}$, 
the right-handed sleptons in the theory need to be lighter than the left-handed ones.  Otherwise the 
invisible decay mode ($\lsptwo\rightarrow\nu\bar\nu\lspone $) takes over 
to suppress this decay mode. On the other hand, $\chonepm$ being 
mostly wino-like does not couple strongly 
to the right-handed sleptons, suppressing the decay 
$\chonepm\rightarrow\lspone e^{\pm}\nu_{e}$ which now can only 
occur via off-shell $W$-boson and the BR can be $\sim$10\% at most as 
can be seen from Fig~\ref{fig:3bd_char_neut}, 
denoted by the red and black points. However, there exist a large 
part of the parameter space, where both $\lsptwo$ and $\chonepm$ may 
have reasonably large BRs into their respective 
three body decay modes. All the lepton generations combined, these BRs 
can be formidable. Later in the collider section, we explore such 
possibilities and find the exclusion limits for this kind of scenarios. 
%%%%%%%%%%%%%%%%%%%%%%%%%%%%%
\subsection{Impact of $(g-2)_{\mu}$}
%%%%%%%%%%%%%%%%%%%%%%%%%%%%%
%\subsection{Impact of $(g-2)_{\mu}$ and Dark Matter}

Existence of DM and the experimentally observed \cite{g-2exp} excess in 
the muon anomalous magnetic moment over the SM prediction \cite{muonrev,g-2sm1,g-2sm2}
remain two of the most robust hints towards BSM physics. 
The BSM contribution to muon anomalous magnetic moment 
(defined as $\Delta a_{\mu}$) has to fit within the deviation 
$\Delta a_{\mu} = (29.3\pm 9.0)\times 10^{-10}$ \cite{muonrev}. 
Efforts have been made to explain this excess in the context of 
various BSM models \cite{ucmc1,ucmc2,Everett:2001tq,nonuni_old,nonuni_new,jcacsm,Nonunivg-2}. 
Within the framework of the MSSM, small slepton and gaugino masses are 
favoured in order to enhance $\Delta a_{\mu}$ to the desired range. 
Thus part of our parameter space is quite relevant from the angle 
of this anomalous experimental result. In general, most of the SUSY 
contribution to $\Delta a_{\mu}$ arises from the chargino-sneutrino loop. 
However, the neutralino-smuon loop can also provide significant 
enhancement depending on the choices of 
left and right-handed smuon mass parameters. In this section,
we use the 1$\sigma$, 2$\sigma$ and 3$\sigma$ allowed ranges of $\Delta a_{\mu}$ in order to 
constrain our parameter space further. 

%%%%%%%%%%%%%%%%%%%%%%%%%%%%%%%
\begin{figure}[h!]
\centering
\hspace{-1cm}
\includegraphics[angle =0, width=0.48\textwidth] {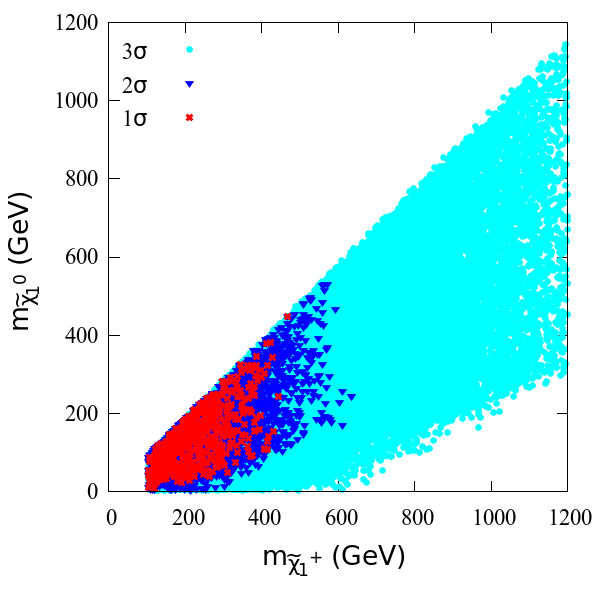}
\includegraphics[angle =0, width=0.48\textwidth] {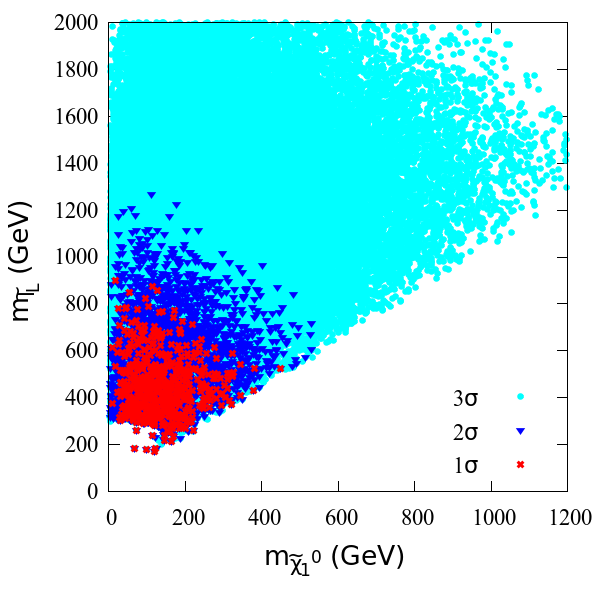}
\caption{Excess in $\Delta a_{\mu}$ obtained at 1$\sigma$, 2$\sigma$ and 3$\sigma$ level shown in the 
$\mchonepm$-$\mlspone$ and $m_{\wt l_L}$-$\mlspone$ plane. The cyan, blue and red points represent 
3$\sigma$, 2$\sigma$ and 1$\sigma$ allowed points respectively.}
\label{fig:g2_gaug}
\end{figure}
%%%%%%%%%%%%%%%%%%%%%%%%%%%%%%

In Fig.~\ref{fig:g2_gaug} (left), we show $\Delta a_{\mu}$ distribution in the 
$\mchonepm$ - $\mlspone$ plane that gives a clear idea of the favoured 
choices of these masses. As evident, 
$\mchonepm > 600$ GeV is not suited well if one intends to take the 
 $\Delta a_{\mu}$ constraint seriously. 
A wide range of the LSP mass is allowed for a particular value of $\mchonepm$, indicating that  
the LSP in these cases can be either bino, wino or a well mixed bino-wino state.   
Another very important factor that goes into the calculation of $\Delta a_{\mu}$ in this framework, 
is the mass range of the sleptons, specially, $m_{\wt\mu_L}$. 
Therefore, in Fig.~\ref{fig:g2_gaug} (right), 
we show $\Delta a_{\mu}$ distribution in the $m_{\wt l_L}$-$\mlspone$ 
plane to give a clear idea about the allowed
ranges of the $m_{\wt\mu_L}$. Clearly, the 1$\sigma$ and 2$\sigma$ allowed 
ranges are at most 900 GeV and 
1250 GeV respectively depending on the choices of the LSP and chargino masses.
%%%%%%%%%%%%%%%%%%%%%%%%%%%%%%
\subsection{Benchmark Points}
%%%%%%%%%%%%%%%%%%%%%%%%%%%%%%
Based on our discussion so far, we have selected a few benchmark points presented in Table.~\ref{tab:bp_elect} below. 
%%%%%%%%%%%%%%%%%%%%%%%%%%%%%%%
\begin{table}[h!]
\begin{center}
\begin{tabular}{||c|c|c|c|c|c|c||} \hline\hline
Parameters  & BP1 & BP2 & BP3 & BP4 & BP5 & BP6 \\ 
\hline\hline 
$M_1$ & 162.4 & 387.4 & 352.4 & 427.3 & 200.9 & 177.1   \\
$M_2$ & 167.2 & 411.2 & 353.3 & 499.3  & 380.9 & 518.3  \\    
$M_3$ & 2000.0 & 2000.0 & 2000.0 & 2000.0 & 2000.0 & 2000.0  \\
$\mu$ & 334.4 & 822.3 & 706.6 & 998.7 & 761.8 & 736.6  \\ 
$\tan\beta$ & 13.2 & 35.7 & 24.7 & 20.6 & 8.9 & 20.5  \\
$M_{l_L}$ & 195.9 & 437.3 & 1840.2 & 785.5 & 1129.8 & 1026.4  \\
$M_{l_R}$ & 929.3 & 1048.9 & 376.5 & 1024.2 & 1055.5 & 586.9  \\
\hline
$m_h$ & 126.1 & 125.7 & 125.6 & 124.7 & 123.7 & 123.5  \\
$m_{\widetilde l_{L}}$ & 201.0 & 439.6 & 1840.7 & 186.8 & 1130.7 & 1027.3  \\
$m_{\widetilde l_{R}}$ & 930.3 & 1049.8 & 379.1 & 1025.1 & 1056.4 & 588.6  \\
$m_{\widetilde \tau_1}$ & 200.9 & 436.4 & 378.8 & 785.1 & 1056.2 & 587.9  \\
$m_{\widetilde \tau_2}$ & 930.3 & 1051.2 & 1840.8 & 1026.3 & 1130.9 & 1028.2  \\
$m_{\widetilde\nu_{L}}$ & 185.5 & 432.7 & 1839.1 & 782.9 & 1128.1 & 1027.7  \\
$m_{\widetilde\chi^0_1}$ & 149.6 & 376.9 & 346.1 & 418.6 & 196.4 & 173.4  \\
$m_{\widetilde\chi^0_2}$ & 167.4 & 426.6 & 370.5 & 519.9 & 395.5 & 531.7  \\
$m_{\widetilde\chi^{\pm}_1}$ & 163.3 & 426.4 & 370.0 & 519.8 & 395.4 & 531.7  \\
\hline
$\rm{BR}(\lsptwo\rightarrow h\lspone)$ & - & - & - & - & 0.94 & 0.83  \\ 
$\rm{BR}(\lsptwo\rightarrow Z\lspone)$ & - & - & - & 1.0 & 0.06 & 0.17  \\
$\rm{BR}(\lsptwo\rightarrow l \bar l\lspone)$ & 1.4$\times 10^{-3}$ & 0.20 & 0.64 & - & - & -  \\
$\rm{BR}(\lsptwo\rightarrow \tau\bar\tau\lspone)$ & 4.8$\times 10^{-3}$ & 0.14 & 0.30 & - & - & -  \\
$\rm{BR}(\lsptwo\rightarrow \nu\bar\nu\lspone)$ & 0.87 & 0.63 & 1.07$\times 10^{-2}$ & - & - & -  \\
$\rm{BR}(\chonepm\rightarrow W^{\pm}\lspone)$ & - & - & - & 1.0 & 1.0 & 1.0  \\ 
$\rm{BR}(\chonepm\rightarrow l\nu\lspone)$ & 0.34 & 0.46 & 0.22 & - & - & -  \\ 
$\rm{BR}(\chonepm\rightarrow \tau\nu\lspone)$ & 0.16 & 0.25 & 0.11 & - & - & -  \\ 
\hline
$\Delta a_{\mu}\times 10^{10}$ & 31.26 & 15.45  & 0.80 & 4.49 & 1.56 & 3.72  \\
$\rm{BR}(b\rightarrow s\gamma)\times 10^{4}$ & 3.37  & 3.42  & 3.40 & 3.38 & 3.36 & 3.39  \\ 
$\rm{BR}(B_s\rightarrow\mu\mu)\times 10^{9}$ & 3.01 & 2.68  & 2.90 & 2.94 & 3.02 & 2.95  \\
%\hline
\hline\hline
\end{tabular}
\caption{Low scale input parameters and the relevant sparticle 
masses along with the values of the relevant 
branching ratios and constraints for some of the chosen benchmark 
points satisfying all the collider, DM and 
low energy constraints discussed in this section. All the mass 
parameters are written in GeV unit.}
\label{tab:bp_elect}
\end{center}
\end{table} 
%%%%%%%%%%%%%%%%%%%%%%%%%%%%%%
BP1, BP2 and BP3 represent part of the parameter space where two-body 
decay modes of the electroweakinos are 
forbidden whereas BP4, BP5 and BP6 represent that where the two-body decays are allowed.  
Different choices of $M_1$, $M_2$, $M_{l_L}$ and $M_{l_R}$ are 
considered to highlight their effect on 
the relevant branching ratios and the experimental constraints. 
For all these benchmark points, squark soft-mass 
parameters are kept fixed at 3 TeV and a large $A_t$ (4 TeV) is considered 
in order to fit the SM-like Higgs mass 
constraint. BP1 and BP2 results in a $\Delta a_{\mu}$ that lie within 
its 2$\sigma$ allowed range. This is mainly because of their light 
chargino and left-smuon mass. It is hard to achieve this kind of enhancement 
in $\Delta a_{\mu}$ once the $M_{l_L}$ parameter starts to increase. 
This feature can be clearly seen from BP3 and BP5.

BP1 has a large mixing between the bino and wino components. As
a result, the $\lsptwo\to\nu\bar\nu\lspone$ decay mode 
dominates over the $\lsptwo\to l\bar l\lspone$ 
mode. In BP2 as this mixing decreases, the invisible decay 
BR starts to fall. In BP3 it becomes negligible 
as a consequence of having large $M_{l_L}$.  
BP4 represents part of the parameter space where the mass gap, 
$\Delta m^0_{\wt\chi}= \mlsptwo - \mlspone$ is greater than 
$m_Z$ but less than $m_h$ and as a result, $\lsptwo$ entirely 
decay into $Z\lspone$. Once the mass gap increases beyond $m_h$, 
the $h\lspone$ decay mode opens up as shown in BP5. In this case, 
the relative branching ratios of these two channels depend upon 
the abundance of the higgsino component in 
both $\lspone$ and $\lsptwo$. BP6 has a relatively smaller 
$\mu$-parameter than BP5 and as a result, has a 
greater branching ratio into $Z\lspone$ mode than in BP5. 
For BP4, BP5 and BP6, $\chonepm$ entirely decays into 
$W^{\pm}\lspone$.

%%%%%%%%%%%%%%%%%%%%%%%%%%%%%%
\section{Collider Analysis}
\label{sec3}
%%%%%%%%%%%%%%%%%%%%%%%%%%%%%% 
ATLAS and CMS collaborations have presented their search results 
 \cite{atlas_ew_3l,atlas_ew_higgs,cmsew2} for direct pair 
production of $\chonepm \lsptwo$ mainly in three types of simplified models: 
(i) \textit {Slepton mediated simplified model}: 
In such scenarios sleptons 
are assumed to be lighter than $\chonepm$ and $\lsptwo$ and the electroweakinos 
decay via slepton to lepton enriched final states \cite{atlas_ew_3l}. 
(ii) \textit {WZ mediated simplified model }:
For these types of models sleptons 
are assumed to be decoupled and the electroweakinos decay via real or virtual 
gauge bosons (BR($\chonepm \ra W \lspone $) = BR($\lsptwo \ra Z \lspone$) = 100\%) 
\cite{atlas_ew_3l}
(iii) \textit {Wh mediated simplified model }: 
In these scenarios also sleptons are decoupled from the 
rest of the electroweak sector and the limits are obtained with 
the assumption that BR($\lsptwo \ra h \lspone$) is 100\% \cite{atlas_ew_higgs}. 
For  \textit {WZ mediated simplified model } ATLAS and CMS has 
looked for trilepton final states with or without 
$\tau$-tagging \cite{atlas_ew_3l,cmsew2}. 
For \textit {Wh mediated simplified model } they have looked for 
final states consisting of an isolated electron or muon with large 
$\met$ associated with any one of the three following possibilities: 
two b-tagged jets ($lb\bar b$ channel), two photons ($l\gamma \gamma$ channel) 
or a second electron or muon of similar electric charge 
($l^\pm l^\pm$ channel) \cite{atlas_ew_higgs}. 
Also ATLAS has presented mass limits on $\mchonepm$ and $\mlsptwo$ 
considering $h \ra \tau\tau, WW, ZZ$ decay modes contributing to trilepton 
final states: 3$l$ ($l = e, \mu$) + 0$\tau$,  2$l$+ 1$\tau$,  1$l$ + 2$\tau$ 
\cite{atlas_ew_3l}. 
Below we briefly discuss about these search analyses \cite{atlas_ew_3l,atlas_ew_higgs} 
used by ATLAS and present our results alongside theirs for validation. Note that, in 
this work, we have only considered the ATLAS analyses.

%%%%%%%%%%%%%%%%%%%%%%%%%%%%%%%%%%
\subsection{Search for Trilepton final states}
\label{sec:3l}
%%%%%%%%%%%%%%%%%%%%%%%%%%%%%%%%%%
%ATLAS collaboration has searched for direct pair production of charginos and 
%neutralinos decaying to a final state with trilepton and missing energy using 
%8 TeV data \cite{atlas_ew_3l}. The limits are interpreted in various simplified 
%models, like  \textit {slepton mediated simplified model} ($\wt\ell_L$-mediated, 
%or $\wt\tau_L$-mediated) \textit {WZ mediated simplified model } and 
%\textit {Wh mediated simplified model }. 
Among all the trilepton channels, the exclusion limit obtained by ATLAS is the 
strongest in $\wt\ell_L$-mediated model, while that from $WZ$- and 
$\wt\tau_L$-mediated are somewhat similar and weaker. \textit {Wh mediated 
simplified model } has the weakest limits. Results are also interpreted in few 
pMSSM scenarios but with fixed values of LSP masses {\cite{atlas_ew_3l}}. 
For validation purpose, we only look into the \textit{$\wt\ell_L$- and $WZ$-mediated 
simplified models}. 
In \textit{$\wt\ell_L$-mediated} models, left handed sleptons and sneutrinos 
are assumed to have mass = ($\mchonemp + \mlsptwo$)/2 and the 
electroweakinos decay either to left handed sleptons or sneutrinos universally. 
In \textit {WZ mediated simplified model }, all the sleptons and sneutrinos 
are assumed to be heavy while the $\chonepm$ and $\lsptwo$ decay 
via real or virtual $W$ and $Z$ respectively with 100 \% BRs.

Events are selected with exactly three tagged leptons (electron, muon or tau) 
with the requirement that one of these tagged leptons must be either electron 
or muon\footnote{Note that here we are not interested in trilepton final states 
comprising of two or more $\tau$-leptons mostly because of their less significant 
results in terms of exclusion limits.}. 
Event reconstruction details like electron, muon, tau and jet 
identification, isolation, overlap removal etc. are followed according to the 
ATLAS analysis as mentioned in Sec. 5 of \cite{atlas_ew_3l}. In this trilepton analysis, 
a veto on b-jet is applied to all signal channels. 
For b-jets, we use the $p_T$ dependent b-tagging efficiencies 
obtained by ATLAS collaboration in Ref. \cite{btagging}. 
For $\tau$-jet identification, we only use the hadronic decay modes. 
We demand that the candidate jets must have $p_{T} >$ 20 GeV and lie 
within $|\eta|<$ 2.5. We also demand that these candidate jets must contain one or three 
charged tracks with $|\eta_{\rm track}| < $ 2.5 and highest track must have $p_{T} > $ 3 GeV. 
Moreover, in order to ensure proper charge track isolation, we put a veto 
on any other charged tracks with $p_{T} > $ 1 GeV inside the candidate jet. 
%%%%%%%%%%%%%%%%%%%%%%%%%%%%%%%
\begin{figure}[h!]
\centering
\hspace{-1cm}
\includegraphics[scale=0.4,angle=270] {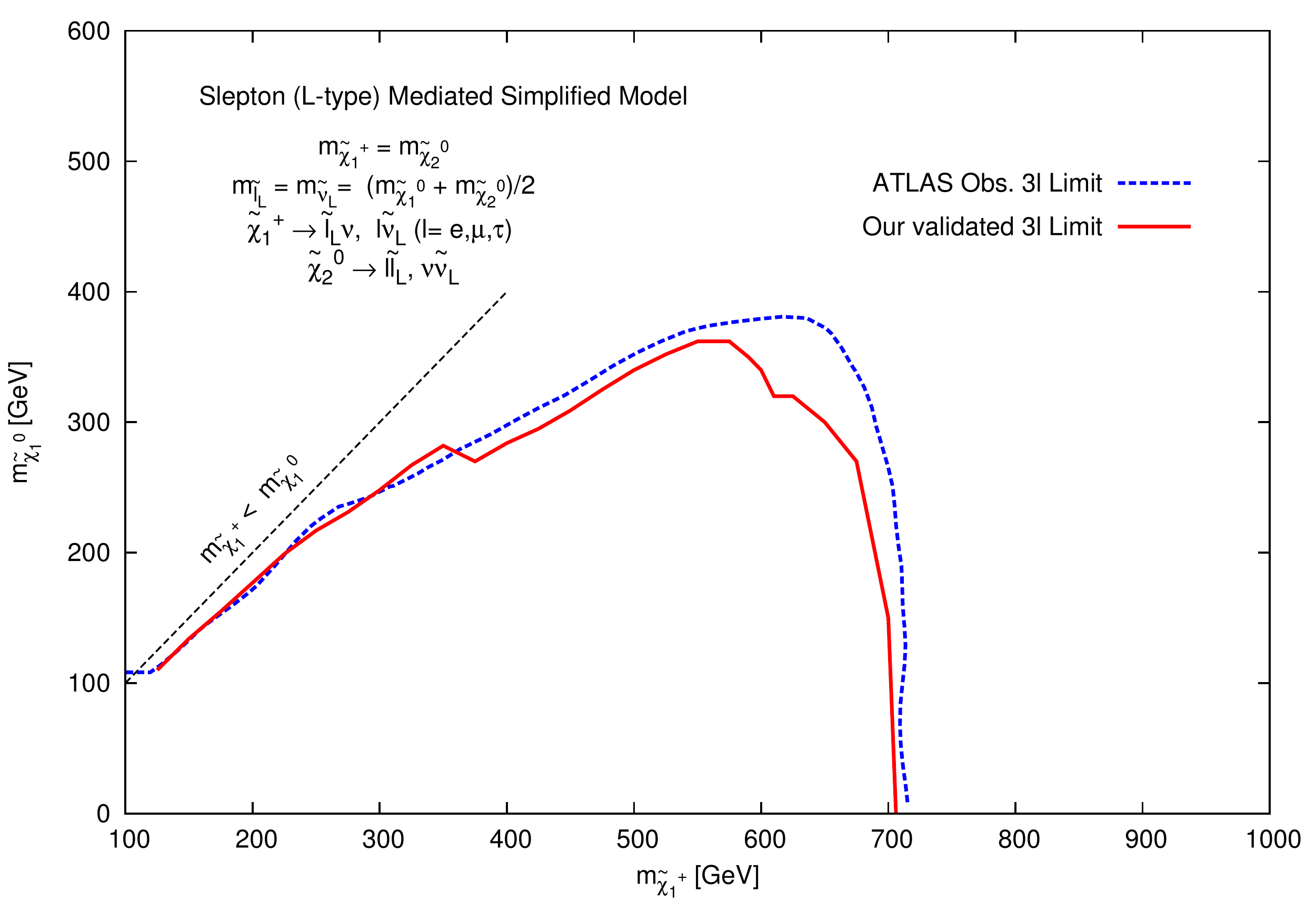}
\caption{ {\it Validation of ATLAS trilepton + $\met$ \cite{atlas_ew_3l} analysis for 
\textit {slepton mediated simplified models}. The blue dotted line corresponds to 
95 \% CL exclusion limits obtained by ATLAS and the solid red line corresponds to our validated results.
}}
\label{3l_slepton}
\end{figure}
%%%%%%%%%%%%%%%%%%%%%%%%%%%%%%%

Depending upon the requirement of number of $\tau$-jets ATLAS has defined 
five signal regions (SR), namely, SR0$\tau$a, SR0$\tau$b, SR1$\tau$, SR2$\tau$a, SR2$\tau$b. 
For implementation of all these signal regions we follow the selection requirements 
as described in Table 3 and Table 4 of Ref.\cite{atlas_ew_3l}. 
Lack of any BSM signal so far in all these channels have resulted in exclusion 
limits presented at the 95\% confidence level (CL), on the number of BSM signal events, $N_{BSM}$, 
for each of the signal regions (SR). These upper limits are presented in Table 7 and Table 8 
of Ref \cite{atlas_ew_3l}. The ATLAS collaboration has translated these obtained upper 
limits on $N_{BSM}$ into exclusion limits in the $\mlspone$ - $\mchonepm$ plane. 
In a similar way, we also have reproduced the exclusion contours obtained by ATLAS 
assuming similar mass relations and branching ratios of the relevant gauginos 
and sleptons. Note that we have also validated the cut-efficiency table provided by 
ATLAS. In order to validate our results we reproduce the exclusion contours using 
PYTHIA (v6.428) \cite{pythia}. 
%We also compute the next to leading order (NLO) 
% chargino-neutralino pair production cross-sections by PROSPINO 2.1 \cite{prospino}. 
% with CTEQ6.6M PDF [175] 
We use the next-to-leading order (NLO) + next-to-leading logarithmic (NLL) 
chargino-neutralino pair production cross-sections given in Ref. \cite{cs_link}, 
which have been calculated for 8 TeV using the resummino code \cite{resummino}.

We observed that for \textit {slepton, $WZ$ or $Wh$ mediated simplified models}, 
SR0$\tau$a and SR0$\tau$b are the most sensitive channels to provide the exclusion 
limits. Henceforth we will not discuss about the rest 
of the three signal regions. Depending upon the invariant mass of same-flavour opposite-sign 
(SFOS) lepton pair ($m_{SFOS}$), which lies closest to the Z boson mass,
SR0$\tau$a signal regions are sliced into five bins and each $m_{SFOS}$ slice 
is further divided into four bins according to the values of $\met$ and 
transverse mass, $m_T$. Here, $m_T$ is constructed with the lepton not 
forming the SFOS pair and $\met$.\footnote{For a summary of these 20 
bins or 20 SR0$\tau$a signal regions, see  
table 4 of \cite{atlas_ew_3l}.} For \textit {slepton mediated models}, 
SR0$\tau$a-bin20 is the most sensitive 
channel for the parameter space with large mass splitting 
between $\chonepm$ and $\lspone$ ($\delta m = $ $\mchonepm - \mlspone$). 
%% corresponds to $m_{SFOS}>$101.2 GeV, $m_T>$120 GeV and $\met>$210 GeV. 
For small $\delta m$, 
low-valued $m_{SFOS}$ SR0$\tau$a bins are 
more effective to probe the relevant parameter space. In Fig.~\ref{3l_slepton}, 
we present the validated results for  \textit {slepton mediated simplified models}.
 The blue dotted line corresponds to 95 \% CL exclusion limits obtained 
by ATLAS and the solid red line corresponds to our validated results adopting 
the ATLAS analysis. From Fig.~\ref{3l_slepton}, it is evident that our validated 
results are in well agreement with that of ATLAS and for low $\mlspone$ ($< 100$ GeV) 
the trilepton channel excludes chargino masses upto 700 GeV. 
%%%%%%%%%%%%%%%%%%%%%%%%%%%%%%%
\begin{figure}[h!]
\centering
\hspace{-1cm}
\includegraphics[scale=0.4,angle=270] {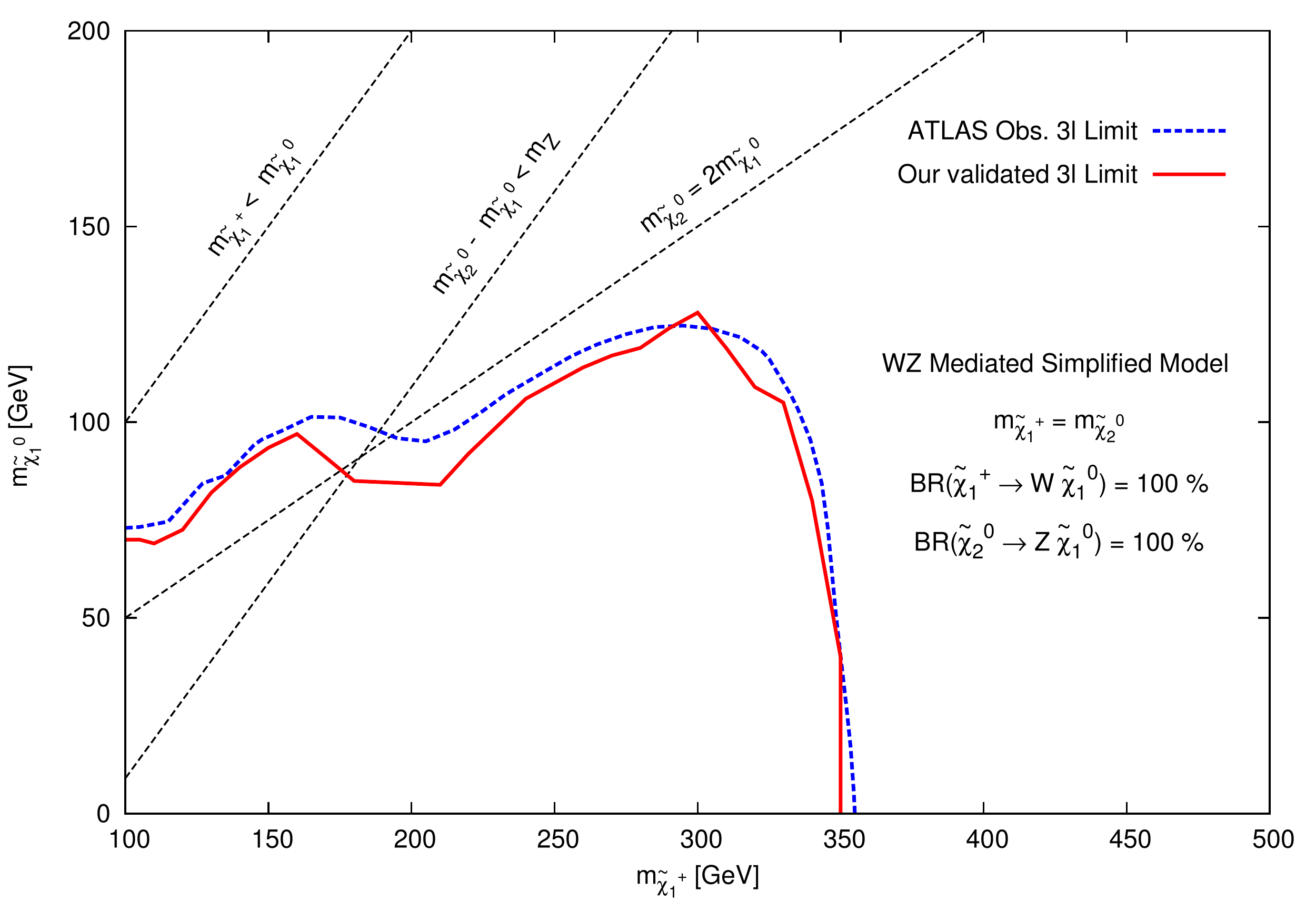}
\caption{Validation of ATLAS trilepton + $\met$ \cite{atlas_ew_3l} analysis for 
\textit {WZ mediated simplified models}. The blue dotted line corresponds to 
95 \% CL exclusion limits obtained by ATLAS and the solid red line corresponds to our validated results.}
\label{3l_wz}
\end{figure}
%%%%%%%%%%%%%%%%%%%%%%%%%%%%%%%

For \textit {WZ mediated simplified models}, the upper limits on $\mchonepm$ is 
relatively weaker ($\mchonepm$ upto 350 GeV are excluded for massless 
$\lspone$). Again for small $\delta m$, SR0$\tau$a-bin01 offers 
the best sensitivity and for large  $\delta m$, 
the exclusion limits are obtained via  SR0$\tau$a-bin16. 
In Fig.~\ref{3l_wz}, we compare the 95\% CL exclusion limit obtained 
by ATLAS (blue dotted line) with the same obtained from our setup 
(red solid line) and they are in good agreement. It may be noted  
that the exclusion line in the regions with $\mlsptwo(\mchonepm) - \mlspone 
< m_Z ~(m_W)$ is obtained from the three body decay of $\lsptwo ~(\chonepm)$
via off-shell gauge bosons.  

%%%%%%%%%%%%%%%%%%%%%%%%%%%%%%%%%%%%%%%%%%%%%%%%%%%%%%%%%%%%%
\subsection{Search for final states with Higgs}
%%%%%%%%%%%%%%%%%%%%%%%%%%%%%%%%%%%%%%%%%%%%%%%
For \textit {Wh mediated simplified models}, where the $\lsptwo$ 
decays into the SM-like Higgs boson and the LSP, 
ATLAS has very recently presented their results 
in two new channels \cite{atlas_ew_higgs}. Along with  large $\met$ and 
an isolated lepton (electron or muon), they have looked for either two b-jets or 
two photons originating from the 125 GeV Higgs. 
Here we have considered these two decay modes of the SM-like Higgs since 
$h\rightarrow b\bar b$ has the largest branching ratio of all the decay modes of 
$h$ and although the $BR(h\rightarrow\gamma\gamma)$ is usually very small, 
the large photon detection efficiency makes this decay mode one of the 
most viable ones in collider studies. We ignore the possible dilepton final state as mentioned 
in the last section, since it gives the weakest limit on the gaugino masses. 
We now discuss about our validation results in these two channels. 

%%%%%%%%%%%%%%%%%%%%%%%%%%%%%%%%%%%%%%%%%%%%%%%%%%%%%%%%%%%%%
\subsubsection{One lepton and two b-jets channel}
%%%%%%%%%%%%%%%%%%%%%%%%%%%%%%%%%%%%%%%%%%%%%%%%%
In this channel, the events are selected with exactly one lepton with $p_T >$ 25 GeV.
We also implement all the isolation criteria used by ATLAS to improve 
the purity of the reconstructed objects and the overlap removal procedure between 
lepton-lepton or lepton-jets summarised in Table 2 of of Ref~\cite{atlas_ew_higgs}. 
In addition to the charged lepton, the events must contain two or three central 
jets. We further ensure that there are exactly two `b-jets' in the final state and 
they must be the two hardest central jets. While tagging the b-jets, we have 
implemented $p_T$ dependent b-tagging efficiency as described in \cite{btagging}. 
Dominant background contributions to this final state are expected to arise 
from $t\bar t$, $W$ + jets and single-top $Wt$ production. A large fraction of 
these SM background events can be suppressed by large missing energy requirement. 
Most dominant $t \bar t$ background can be further  suppressed by a suitable cut 
on the contranverse mass, $m_{CT}$ \cite{Tovey:2008ui,Polesello:2009rn}  
of the b-jet pair, defined as 
\beq
m_{CT} = \sqrt{(E_T^{b_1}+E_T^{b_2})^2 - |\vec{p_T}^{b_1}-\vec{p_T}^{b_2}|^2}, \nonumber
\eeq
where, $E_T^{b_i}$ and $\vec{p_T}^{b_i}$ are the transverse energy and momentum of the $i$-th $b$-jet.
Finally, depending upon the values of $W$ transverse mass 
($m_T^W  = \sqrt{2E_T^{l}{\met} - 2\vec{p_T}^{l}.\vec{\PMET}}$, 
where $E_T^{l}$ and $\vec{p_T}^{l}$ are the transverse energy and 
momentum of the isolated lepton) ATLAS collaboration has defined 
two signal region: SR$lbb$-1 which is 
sensitive to low mass splitting between 
$\mlsptwo$ and $m_h$ and SR$lbb$-2 which is sensitive to large mass 
splitting between $\mlsptwo$ and $m_h$. 
Details of the selection requirement of SR$lbb$-1 and SR$lbb$-2 
signal regions are enlisted in Table~\ref{tab1l2b}. 
In absence of any excess in this channel, ATLAS collaboration 
has derived an upper limit on the number of BSM 
signal events \cite{atlas_ew_higgs} which are quoted  in the 
 last row of Table~\ref{tab1l2b}.

%%%%%%%%%%%%%%%%%%%%%%%%%%%%%%%%%%%%%%
\begin{table}[!h]
\begin{center}\
\begin{tabular}{||c||c|c||}
\hline
\hline
	 			&SR$lbb$-1   		&  SR$lbb$-2 	\\
\hline
$N_{lepton}$		& 1	 				&	1	\\
\hline
$N_{jet}$		& 2-3 				&	2-3	\\
\hline
$N_{b-jet}$		& 2	 				&	2	\\
\hline
$\met >$ (GeV)		& 100	 		&	100	\\
\hline
$m_{CT}$ (GeV)		& $>$160	 		&	$>$160	\\
\hline
$m_{T}^W$		& 100 - 130	 				&	$>$130	\\
\hline
Observed upper limits on	 & & \\
$N_{BSM}$ (95 $\%$ CL)	& 5.3&5.5 \\
\hline
\hline
       \end{tabular}\
       \end{center}
           \caption{Selection requirements and 95 $\%$ upper limit on the number of 
events at 8 TeV with $\lum$ = 20.3 $\ifb$ for SR$lbb$-1 and SR$lbb$-2 signal regions   }
\label{tab1l2b}
\end{table}
%%%%%%%%%%%%%%%%%%%%%%%%%%%%%%%%%%%%%%%%%%%%%%%%% 

%%%%%%%%%%%%%%%%%%%%%%%%%%%%%%%
\begin{figure}[h!]
\centering
\hspace{-1cm}
\includegraphics[scale=0.4,angle=270] {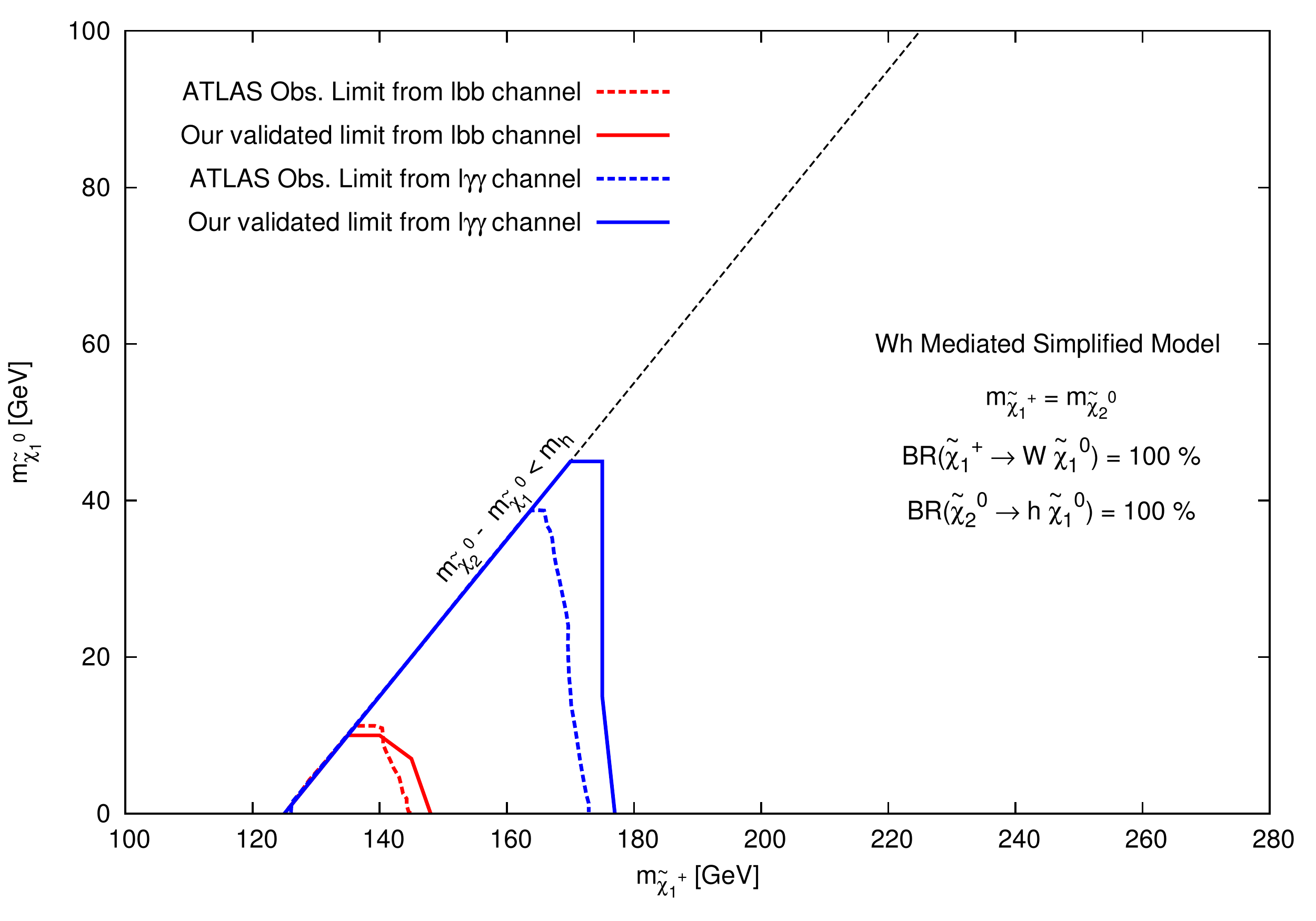}
\caption{Validation of ATLAS $lbb$ + $\met$ and $l\gamma\gamma$ + $\met$ \cite{atlas_ew_3l} analyses for 
\textit {Wh mediated simplified models}. The blue dotted and solid lines correspond to the 95 \% CL exclusion limits obtained by ATLAS and our validated results respectively corresponding to 
$l\gamma\gamma$ + $\met$ final state. Similarly, the red dotted and solid lines represent the experimental 
bound and our validated results corresponding to $lbb$ + $\met$ final state.}
\label{wh_val}
\end{figure}
%%%%%%%%%%%%%%%%%%%%%%%%%%%%%%%

In Fig.~\ref{wh_val}, we reproduce the the exclusion contour in 
$\mlspone$ - $\mchonepm$ plane obtained by ATLAS with 95\% CL using 
8 TeV data \cite{atlas_ew_higgs}. In a similar manner described in 
Sec. \ref{sec:3l} we validate our simulations. 
% The ATLAS collaboration has translated these obtained upper 
% limits on $S$ into exclusion limits in the $\mlspone$ - $\mchonepm$ plane. 
%We present the results obtained by us and the ATLAS collaboration in Fig.~\ref{wh_val} for comparison. 
The red dotted line corresponds to ATLAS and the red thick line corresponds to our validated results. 
From Fig.~\ref{wh_val}, it is clear that the exclusion line obtained 
by us is pretty similar to that of ATLAS.

%%%%%%%%%%%%%%%%%%%%%%%%%%%%%%%%%%%%%%%%%%%%%%%%%%%%%%%%%%
\subsubsection{One lepton and two photons channel}
%%%%%%%%%%%%%%%%%%%%%%%%%%%%%%%%%%%%%%%%%%%%%%%%%
As already mentioned, the other channel that ATLAS collaboration has also 
considered while looking for \textit {Wh mediated simplified models}, 
consists of events with exactly one charged-lepton and two photons in 
the final state. The events are selected either with single-lepton or 
with di-photon trigger. For the single-lepton trigger, events are 
required to have isolated leptons with with $p_T^{l}>$ 25 GeV 
($l = e$ or $\mu$) and two leading photons with $p_T^{\gamma_1}>$ 40 GeV 
(leading) and $p_T^{\gamma_2}>$ 20 GeV (subleading). For diphoton trigger, 
the event selection requires 
$p_T$ thresholds of 15 (10) GeV  for electrons (muons), and 40 (27) GeV 
for the leading (sub-leading) photon. 
The most dominant background contributions to this channel come from
 multi-jet and $Z\gamma$ production, 
where the leptons or jets may be mistagged as photons. 
An optimum $\met > 40$ GeV cut helps to 
suppress these backgrounds.  While reconstructing the $\vec{p_T}$ of the 
 $W \ra l \nu$ system, it is assumed that $\vec{p_T}$ of the 
neutrino is same as $\vec{\PMET}$ and that it is back to back with the 
$h \ra \gamma \gamma$ candidate 
($\delta \phi (W,h) > 2.25$). These events are divided into two SRs 
 (SR$\ell\gamma\gamma$-1 and SR$\ell\gamma\gamma$-2) based on the values 
 of the transverse mass of the $W\gamma_i$ system, 
$m_T^{W\gamma_i}$, defined as
\beq
m_T^{W\gamma_i} = \sqrt{(m_T^{W})^2 + 2E_T^{W}E_T^{\gamma_i} - 2\vec{p_T}^W.\vec{p_T}^{\gamma_i}}, \nonumber
\eeq
where, $m_T^W$, $E_T^W$ and $\vec{p_T}^W$ are respectively 
the transverse mass, energy and momentum of 
$W$ and $E_T^{\gamma_i}$ and $\vec{p_T}^{\gamma_i}$ are the 
transverse energy and momentum of the $i$-th 
photon. All the cuts \cite{atlas_ew_higgs} applied for these two 
 different signal regions are listed in 
Table~\ref{tab1l2g}. 
%%%%%%%%%%%%%%%%%%%%%%%%%%%%%%%%%%%%%%
\begin{table}[!htb]
\begin{center}\
\begin{tabular}{||c||c|c||}
\hline
\hline
	 			&SR$l\gamma \gamma$-1   		&  SR$l\gamma \gamma$-2 	\\
\hline
$N_{lepton}$		& 1	 				&	1	\\
\hline
$N_{\gamma}$		& 2-3 				&	2-3	\\
\hline
$\met >$ (GeV)		& 40	 		&	40	\\
\hline
$\delta\phi(W,h)$ (GeV)		& $>$160	 		&	$>$160	\\
\hline
$m_{\gamma \gamma}$ range (GeV)		& [100,160]	 		&[100,160]	\\
\hline
$m_{T}^{W \gamma_1}$		& $>$150 				&	$<$150	\\
						& and	 				&	or	\\
$m_{T}^{W\gamma_2}$		& $>$80 				&	$<$80	\\
\hline
Observed upper limits on	 & & \\
$N_{BSM}$ (95 $\%$ CL)	& 3.6&7.0 \\
\hline
\hline
       \end{tabular}\
       \end{center}
           \caption{Selection requirements and 95 $\%$ upper limit on the number of 
events at 8 TeV with $\lum$ = 20.3 $\ifb$ for SR$l\gamma\gamma$-1 and SR$l\gamma\gamma$-2 signal regions   }
\label{tab1l2g}
\end{table}
%%%%%%%%%%%%%%%%%%%%%%%%%%%%%%%%%%%%%%
Following the same procedure (discussed in previous subsection), we have validated 
the exclusion limit obtained by ATLAS for Wh simplified scenarios. We present 
the results in Fig.~\ref{wh_val}. As evident, our validated results (solid blue line) are in 
good agreement with the ATLAS exclusion line (blue dotted line).
%%%%%%%%%%%%%%%%%%%%%%%%%%%%%%%%%%%%%%%%%

%%  For Wh-mediated simplided models --- to draw the exclusion line they  have statistically combined (!!!!!!) the four signal regions SR0$\tau$a, SR0$\tau$b, SR1$\tau$, SR2$\tau$b.  However, each single channel gives much weaker limit than the combined limit presented by ATLAS. Among the four signal regions,  we observe that SR0$\tau$b is the best sensitive signal region. 
%%  In this work, we have not combined (statistically) any signal regions.... In fig we present only the limit obtained by SR0$\tau$b signal region. ----  20 GeV weaker !!!!!!!!!!! Also mention that 1l2$\gamma$ channel is more stronger than 3l channel. ----- May be we should not give any plot ---- !! 

%%%%%%%%%%%%%%%%%%%%%%%%%%%%%%%%%%%%%%%%%%%%%%%%%%%
\subsection{Revisiting the exclusion limits with varying branching ratios}
%%%%%%%%%%%%%%%%%%%%%%%%%%%%%%%%%%%%%%%%%%%%%%%%%%%
In this section, we revisit the aforementioned search channels varying the 
relevant branching ratios into a particular 
decay mode to study their impact on the existing exclusion limits provided 
by the experimental collaborations as
discussed in the previous subsection. Quite obviously, the limits are 
expected to get weaker if one considers 
shared decay modes of the electroweakinos instead of assuming their 
wholesome decay into one particular decay mode. 
For example, in \textit {Wh } and \textit {WZ mediated simplified models} 
it is assumed that the branching ratios, $\lsptwo \to h \lspone $ and $Z \lspone$ are 
100\% respectively. But in Sec.~\ref{sec2} we have shown that in the allowed kinematic 
region (see Fig.~\ref{fig:neut2-2body}) these two decay modes can compete with each 
other. Our aim is to assess how much one may expect the exclusion limits to change under such more realistic situations. 
%%%%%%%%%%%%%%%%%%%%%%%%%%%%%%%
\begin{figure}[h!]
\centering
\hspace{-1cm}
\includegraphics[scale=0.4,angle=270] {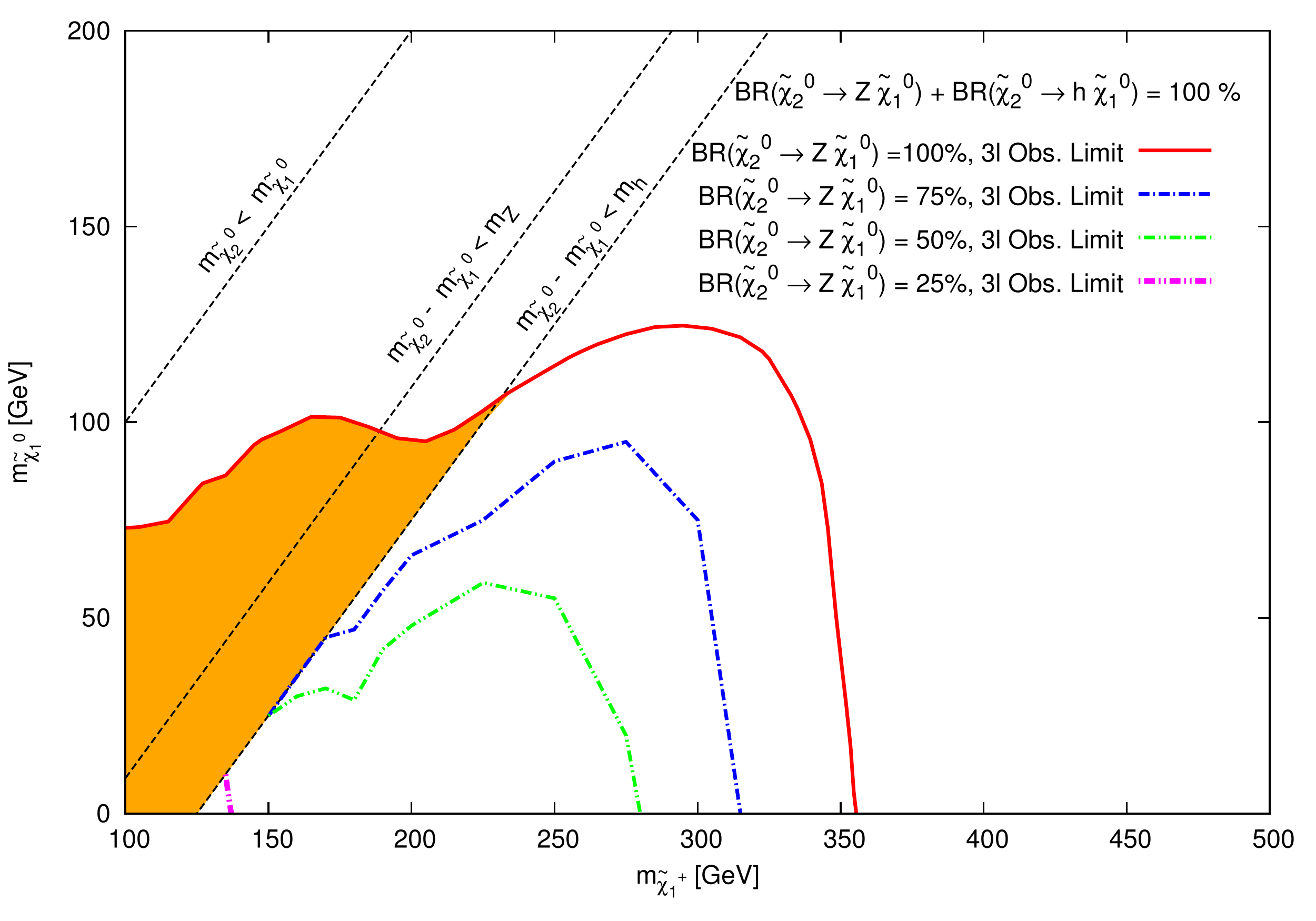}
\caption{The exclusion lines shown in $\mlspone$ - $\mchonepm$ mass plane for 
trilepton final state with different choices of BR($\lsptwo \to Z \lspone $). 
The red line represents the present experimental bound, whereas the blue, 
green and magenta dotted lines present our results 
obtained assuming BR($\lsptwo \to Z \lspone $) = 75\%, 50\% and 25\% respectively. 
The black dotted lines separate various 
kinematic regions where one particular decay mode ceases to exist and another 
opens up. The yellow shaded region represents the parameter space where 
either $\lsptwo \to h \lspone $ or $\lsptwo \to Z \lspone $ or both these 
decay modes are forbidden.}
\label{wzh_3l}
\end{figure}
%%%%%%%%%%%%%%%%%%%%%%%%%%%%%%%
We also consider the scenario where electroweakinos are kinematically 
forbidden to decay into real $W/Z$. 
Under such circumstances, we observe that there exist a large part 
of the parameter space where the charged sleptons 
are heavier than $\lsptwo$ and $\chonepm$, but not so heavy so 
that they may be considered to be decoupled from the rest 
of the spectrum, the off-shell decays of the gauginos provide 
exclusion limits much stronger than that obtained from the 
usual two-body decay modes.

For the sake of simplicity, first we choose a simplified model 
where the two body decay modes of $\lsptwo$ (wino-like) are varied. 
We assume that BR($\lsptwo \to Z \lspone $) + BR($\lsptwo \to h \lspone $) = 100\%, 
which is true in general if the sleptons are heavier than $\lsptwo$. 
Then for illustrative purpose, we derive the revised limit for 
BR($\lsptwo \to Z \lspone $) = 75\%, 50\% and 25\% respectively. 
In Fig.~\ref{wzh_3l} we present the effect of decreasing 
$Z \lspone $ branching ratios over the trilepton final states.

The red solid line represents the exclusion line obtained for 
\textit {WZ mediated simplified models} with 100\% BR to 
WZ\footnote{The sleptons being heavier, the only two-body decay 
mode available to $\chonepm$ is $W^{\pm}\lspone$.} from 
trilepton modes (the same line from Fig.~\ref{3l_wz}), 
while the blue, green and magenta dotted line represent the 
exclusion contours for BR($\lsptwo \to Z \lspone $) = 75\%, 50\% and 25\% 
respectively. It is clear that the limits reduce drastically 
due to enhancement of $\lsptwo \to h \lspone $ decay. 
The black dotted lines separate different kinematical regions 
of interest as indicated in the plot. The yellow shaded region 
is the kinematical region where the trilepton bound is obtained 
from solely $\lsptwo \to Z \lspone$ decay mode (if 
$\mlsptwo - \mlspone < m_h$) or from three-body decays of 
$\lsptwo$ (if $\mlsptwo - \mlspone < m_Z$) via off-shell Z.
Note that the trilepton limit almost vanish for BR($\lsptwo \to Z \lspone $) = 25\%. 

%%%%%%%%%%%%%%%%%%%%%%%%%%%%%%%
\begin{figure}[h!]
\centering
\hspace{-1cm}
\includegraphics[scale=0.4,angle=270] {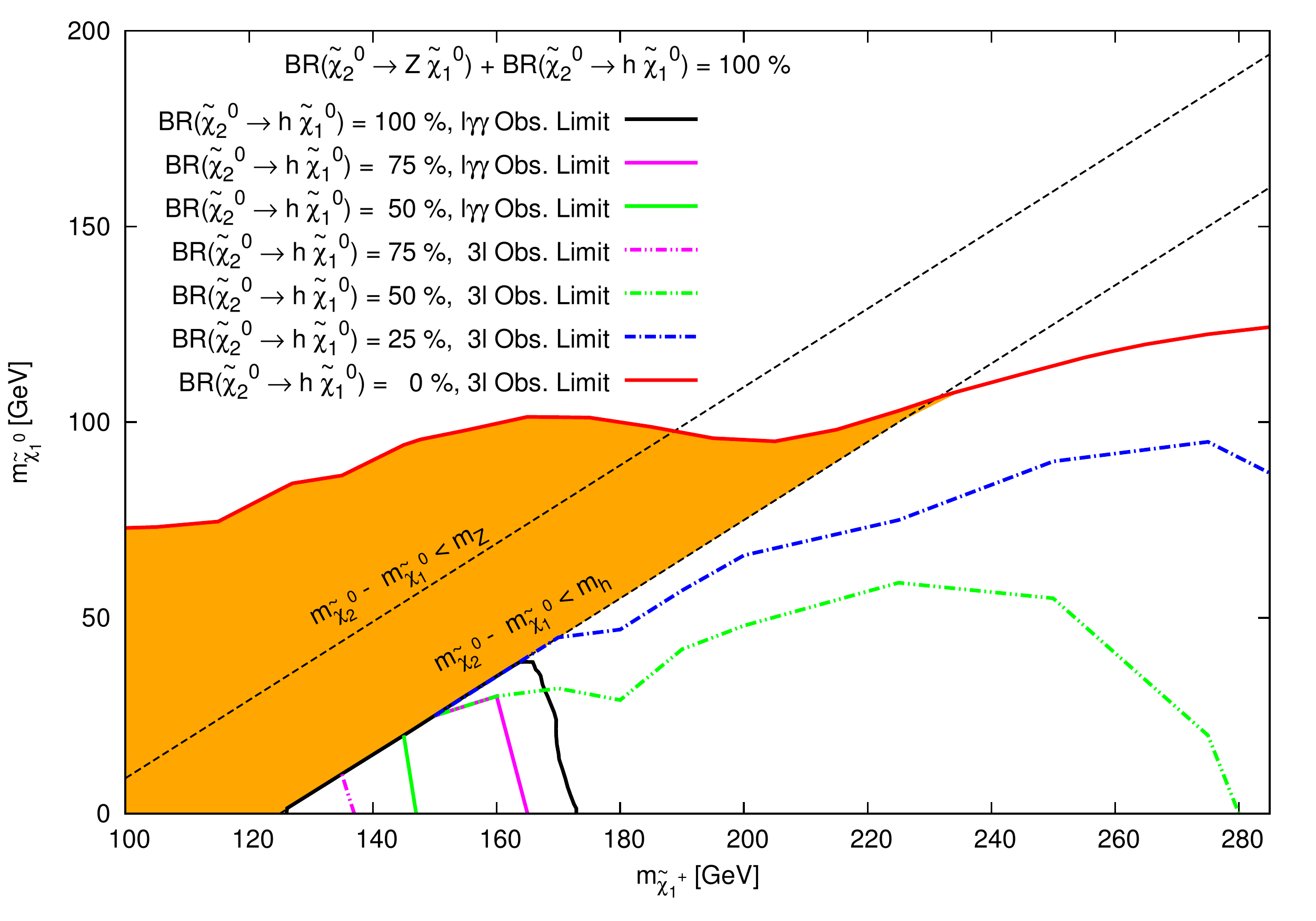}
\caption{The exclusion lines shown in $\mlspone$ - $\mchonepm$ mass 
plane for $l\gamma\gamma$ and trilepton final states 
with different choices of BR($\lsptwo \to h \lspone $). 
The solid black line represents the present experimental bound 
for $l\gamma\gamma$ final state, whereas the solid magenta 
and solid green lines present our results obtained assuming 
BR($\lsptwo \to h \lspone $) = 75\% and 50\% respectively. 
The solid red, dotted blue, dotted green and dotted magenta 
colored lines are the same as shown in Fig.~\ref{wzh_3l}.}
\label{wzh_lgg}
\end{figure}
%%%%%%%%%%%%%%%%%%%%%%%%%%%%%%%
Next, we concentrate on the decay mode $\lsptwo \to h \lspone $. 
In Fig.~\ref{wzh_lgg} we present the behaviour 
of $1l+2\gamma$ channel where the $h$ decays into two 
photons\footnote{$1l+2b$ channel provides much weaker limit.}.

The solid magenta and green lines represent the exclusion 
limits obtained for BR($\lsptwo \to h \lspone $) = 75\% and 50\% whereas 
the solid black line which is the experimental bound obtained 
assuming this particular decay BR to be 100\%.  
However, this channel provides much weaker limits compared to 
the trilepton channel. To showcase this, we have also 
shown the exclusion lines obtained from the trilepton final 
states alongside that from $l\gamma\gamma$ final state. 
The blue and green dotted lines shown in this plot are the 
same as shown in Fig.~\ref{wzh_3l}. Evidently, the trilepton 
bounds are stronger unless the BR($\lsptwo \to Z \lspone $) 
is close to 25\% or smaller. The black dotted lines and the 
yellow shaded region shown in this plot are the same as in Fig.~\ref{wzh_3l}.

%%%%%%%%%%%%%%%%%%%%%%%%%%%%%%%
\begin{figure}[h!]
\centering
\hspace{-1cm}
\includegraphics[scale=0.4,angle=270] {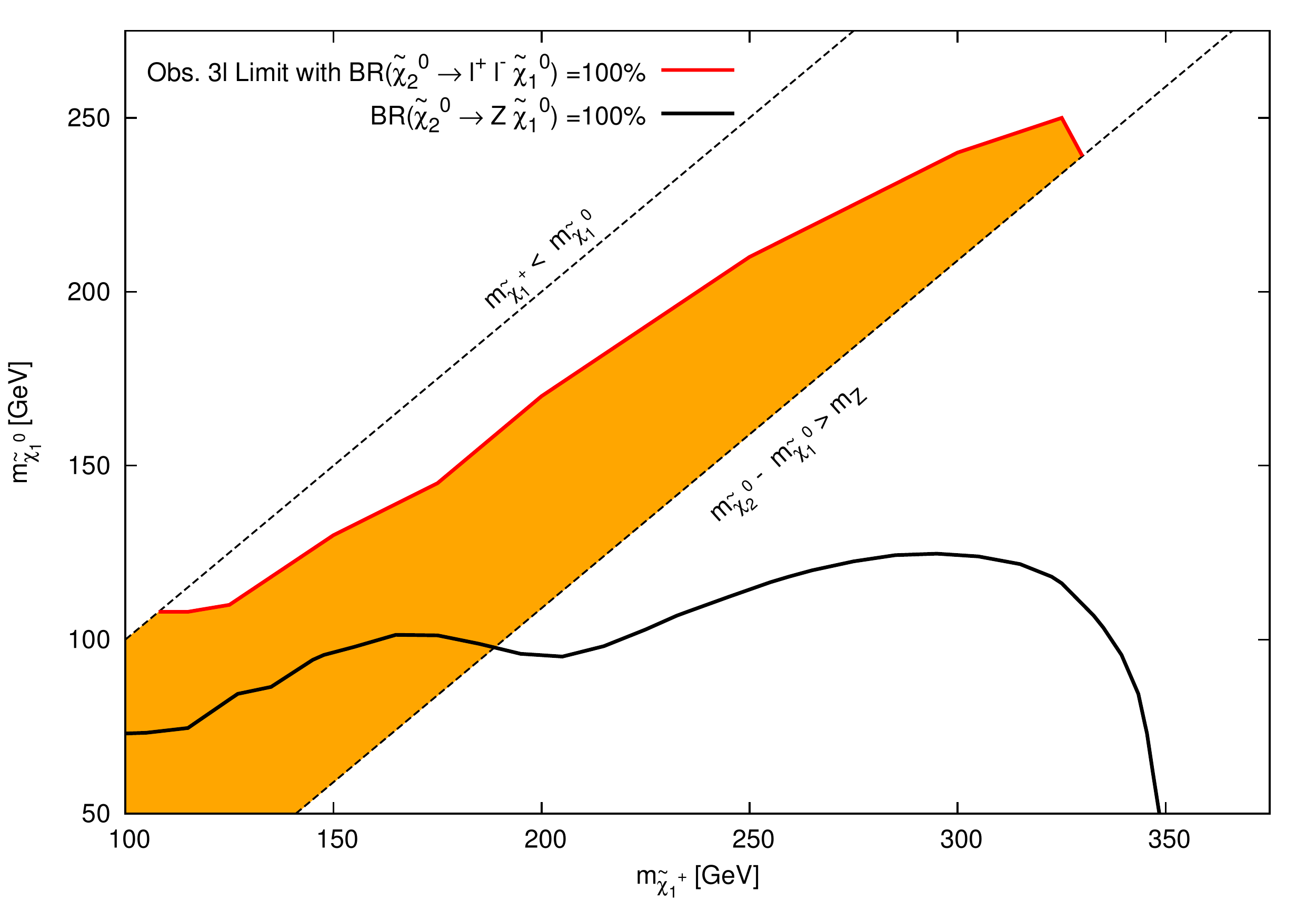}
\caption{The exclusion lines shown in $\mlspone$ - $\mchonepm$ mass plane for trilepton final state assuming 
BR($\lsptwo \to l\bar l \lspone$) = 100\% and BR($\chonepm\rightarrow\lspone l^{\pm}\nu_{l}$) = 30$\%$ where 
$l=e, \mu, \tau$. The solid black line represents the present 
experimental bound for trilepton final state, whereas 
the yellow shaded region below the red line shows the obtained exclusion region from trilepton data for 
such off-shell decays of the electroweakinos.}

% the red line present our result obtained assuming 
% The yellow shaded region shows the obtained exclusion region from trilepton data for such off-shell decays 
% of the electroweakinos.}
\label{fig_3body}
\end{figure}
%%%%%%%%%%%%%%%%%%%%%%%%%%%%%%%
Finally, we consider the scenario where none of the two-body decays are kinematically allowed for both $\lsptwo$ and 
$\chonepm$. Under these circumstances, $\lsptwo$ and $\chonepm$ may decay via off-shell sleptons\footnote{Electroweakinos 
decaying into on-shell sleptons have been studied experimentally. We are not considering that scenario here.} and still 
give rise to the trilepton signal provided the sleptons are not too heavy as can be understood from Fig.~\ref{fig:neut2-3body} 
and \ref{fig:char-3body}. To showcase this, we construct a simplified model motivated by Fig.~\ref{fig:3bd_char_neut}.  
In this simplified model, $\lsptwo$ decays entirely into the three charged lepton pairs universally associated with $\lspone$, 
i.e, BR($\lsptwo\to l \bar l \lspone$) = 100\% where $l=e, \mu, \tau$. $\chonepm$ also decays universally into all its three-body 
leptonic modes via an off-shell $W$-boson. However, as indicated in Fig.~\ref{fig:3bd_char_neut}, for a large 
BR($\lsptwo\to l \bar l \lspone$), the other relevant 3-body BR($\chonepm\rightarrow\lspone l^{\pm}\nu_{l}$) remains suppressed. 
$\chonepm$ now decays via an off-shell W-boson and its combined leptonic 3-body BR can be at most 30$\%$
Under this scenario, the new exclusion limits obtained from trilepton channel are shown in Fig.~\ref{fig_3body}.  
The black solid line represents the exclusion line when $\lsptwo$ decays only via real or virtual Z. In the region 
between two dotted black line the two body decay mode via real Z is kinematically not allowed. Now in presence 
of light slepton we estimate that the whole yellow shaded region under the red solid line are excluded from 
trilepton data. As evident, this limit is much stronger than the conventional one (black line) within the 
region enclosed by the two dotted black lines. 
%%%%%%%%%%%%%%%%%%%%%%%%%%%%%%%
\begin{figure}[h!]
\centering
\hspace{-1cm}
\includegraphics[scale=0.4] {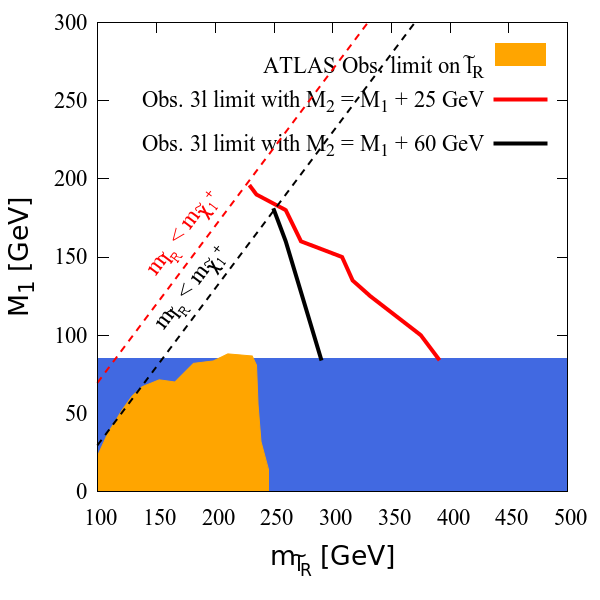}
\caption{The exclusion lines shown in $M_1$ - $m_{\widetilde l_R}$ mass plane for trilepton final state assuming 
assuming varying branching ratio. The orange region represents the experimental limit on $m_{\widetilde l_R}$ 
derived from direct production of the sleptons provided the $m_{\widetilde l_L}$ are decoupled from the rest of the spectrum. 
The red and black solid lines correspond to the exclusion lines obtained for $M_2=M_1$+25 GeV and $M_2=M_1$+60 GeV 
respectively. The dotted red and black lines correspond to $m_{\widetilde l_R} < \mchonepm$ 
for the two scenarios: $M_2=M_1$+25 GeV and $M_2=M_1$+60 GeV.  }
\label{fig_3body_soft}
\end{figure}
%%%%%%%%%%%%%%%%%%%%%%%%%%%%%%%

In Fig.~\ref{fig_3body_soft} we show the trilepton exclusion line obtained from off-shell slepton decays keeping 
$M_2$ and $M_1$ at two specific intervals in the $M_1$-$m_{\widetilde l_R}$ mass plane for clarity. Note that, unlike 
Fig.~\ref{fig_3body}, here we do not take a fixed branching ratio of $\lsptwo$ and $\chonepm$. 
Instead, we present a more generalised scenario, where this BR may vary freely to give an idea how the exclusion limit
applies to the few parameters involved (here the choices being $M_1$ and $m_{\widetilde l_R}$) for such decay modes. 
In Fig.~\ref{fig_3body_soft} the choice of $M_2=M_1+X$ ($X$ chosen such that $\mlsptwo$ - $\mlspone < m_Z$) automatically 
puts a lower limit on the LSP mass. 
However, the shaded blue region is the excluded neutralino mass region irrespective of our choice of $X$ since even in the absence 
of large off-shell slepton decay BR, the contribution arising from off-shell Z-boson decay rules out this part of the parameter space. 
The dotted red and black lines correspond to $m_{\widetilde l_R} < \mchonepm$ 
depending upon the choices of $X$. As expected, $M_2=M_1$+60 case gives a weaker exclusion limit because of heavier $\lsptwo$ and $\chonepm$.    
As $X$ goes down, the exclusion limit on $m_{\widetilde l_R}$ strengthens. However, we chose not to go below 25 GeV, since beyond this limit, 
the parameter space is highly compressed and the final state leptons are likely to escape detection. This part of the parameter space is 
clearly visible in Fig.~\ref{fig_3body} in between the solid red and dotted black line representing $\mchonepm < \mlspone$.   

Note that, these 
limits are comparable to that obtained from $l_L$-mediated simplified models (see Fig.~\ref{3l_slepton}). This 
implies that even heavier sleptons may result in similar exclusion limits and thereby emphasises 
the need to probe these off-shell decay modes of the electroweakinos more carefully.  
%%%%%%%%%%%%%%%%%%%%%%%%%%%%%%  
\section{Summary and Conclusions}
\label{sec4}
%%%%%%%%%%%%%%%%%%%%%%%%%%%%%%
In the absence of any significant results towards the discovery of new physics beyond 
the SM, the experimental collaborations have extensively studied the obtained data 
so far to put exclusion limits on the possible BSM scenarios. These experimental limits 
act as guiding lights toward our quest of BSM physics. However, one has to choose 
these mass limits judiciously as a lot of simplified assumptions are made in order 
to obtain such limits. In this work, we have revisited the exclusion limits on the 
gaugino masses derived from the Run-I data at the LHC. The experimental collaborations 
have looked into various final states comprised of leptons, jets and missing energy 
in order to put the exclusion limits in the $\mlspone$ - $\mchonepm$ plane. 
While deriving these limits, they work with some simplified models where $\lsptwo$ 
and $\chonepm$ decay entirely into one of their possible decay modes 
which although true for some part of the parameter space, often do not present the whole picture. In 
realistic scenarios, the various decay modes may compete with each other and the exclusion limits are 
expected to change significantly. Our aim was to assess how much deviation of these exclusion limits 
one may expect for such scenarios. For that purpose, we have scanned the pMSSM parameter space to 
find our region of interest where combinations of different decay modes of $\lsptwo$ and $\chonepm$ 
may also give rise to the similar final states studied by the experimental 
collaborations. We show some 
representative benchmark points obtained from our scan satisfying all the 
relevant experimental constraints 
to show the interplay between the branching ratios of the various available 
decay modes of $\lsptwo$ and $\chonepm$. 
We validate our results with those of ATLAS using their assumptions before proceeding to explore the 
effects of the interplay of the different branching ratios. The obtained results are presented in 
$\mlspone$ - $\mchonepm$ plane along with the existing 
experimental results to showcase the significant deviations. 
We also observe that in the absence of the two-body decay modes, 
the three-body decays of $\lsptwo$ and $\chonepm$ 
via off-shell sleptons can also give rise to the trilepton final 
states and the exclusion limits obtained are far more 
severe than that obtained from off-shell gauge boson decays.   
%%%%%%%%%%%%%%%%%%%%%%%%%%%%%%
\section*{Acknowledgments} 
%%%%%%%%%%%%%%%%%%%%%%%%%%%%%%
The work of AC is supported by the Lancaster-Manchester-Sheffield 
Consortium for Fundamental Physics under STFC Grant No. ST/L000520/1. 
The work of SM is partially supported by funding available from the 
Department of Atomic Energy, Government of India, for the Regional
Centre for Accelerator-based Particle Physics (RECAPP), 
Harish-Chandra Research Institute. 
The authors also acknowledge useful discussions with 
Amitava Datta, Utpal Chattopadhyay, Sujoy Poddar and Manimala Chakraborti.  
%%%%%%%%%%%%%%%%%%%%%%%%%%%%%%
%----------------------------------
\input{references.tex}

%----------------------------------
\end{document} 

%%%%%%%%%%%%%%%%%%%%%%%%%%%%%%%%%%%%%%%%%%%%%%%

%% file: references.tex
%\bibliographystyle{unsrt}